\documentclass[aps,prl,twocolumn,showpacs,superscriptaddress,floatfix]{revtex4-2} %twocolumn,
\usepackage{graphicx}% Include figure files
\usepackage{dcolumn}% Align table columns on decimal point
\usepackage{bm}% bold math
\usepackage{hyperref}
\usepackage[utf8]{inputenc}
\usepackage[english]{babel}
\usepackage{units}
\usepackage{mathtools}
\hypersetup{colorlinks=true,citecolor={blue},linkcolor={blue},urlcolor={blue}}
\usepackage{float}
\usepackage{amsmath}
\usepackage{empheq}
\usepackage{mathrsfs}
\usepackage{amssymb}
\usepackage{soul}
\usepackage{textcomp}
\usepackage{placeins}
\usepackage{xcolor}
\usepackage[T1]{fontenc}
\usepackage{tipa}
\setstcolor{red}

\begin{document}

\title{Non-local Stabilization and Synchronization of Larmor Precession in Trapped Polariton Condensate Dimer}

\author{S. Baryshev}
\email{stepan.baryshev@skoltech.ru}
\address{Skolkovo Institute of Science and Technology, Moscow, Territory of innovation center “Skolkovo”, Bolshoy Boulevard 30, bld. 1, 121205, Russian Federation.}

\author{A. Yulin}
\address{School of Physics and Engineering, ITMO University, St. Petersburg 197101, Russian Federation.}

\author{I. Gnusov}
\address{Skolkovo Institute of Science and Technology, Moscow, Territory of innovation center “Skolkovo”, Bolshoy Boulevard 30, bld. 1, 121205, Russian Federation.}

\author{S. Alyatkin}
\address{Skolkovo Institute of Science and Technology, Moscow, Territory of innovation center “Skolkovo”, Bolshoy Boulevard 30, bld. 1, 121205, Russian Federation.}

\author{P. G. Lagoudakis}
\address{Skolkovo Institute of Science and Technology, Moscow, Territory of innovation center “Skolkovo”, Bolshoy Boulevard 30, bld. 1, 121205, Russian Federation.}

\begin{abstract}
We demonstrate mutual synchronization of Larmor precessions between spatially separated, optically trapped polariton condensates. We identify the cross-condensate correlation $g^{(2)}_{1,2}(\tau)$  as a sensitive metric for condensate coupling. Synchronization of precessions in a condensate dimer is mediated by polariton outflow and resonantly depends on spectral overlap and precession winding. All-optical stirring of one site non-locally increases the pseudo-spin dephasing time of its static neighbor, prolonging phase-locking in a dimer by an order of magnitude. Our work establishes a scalable platform for studying non-equilibrium time-periodically driven systems and for high-speed spinoptronic networks. 

\end{abstract}

\maketitle
Synchronization is a universal phenomenon spanning biological, chemical, and physical systems~\cite{Pikovsky_Rosenblum_Kurths_2001}. From Huygens’ seminal observations of phase-locking in pendulum clocks to the coordinated firing of neurons in the mammalian brain, the spontaneous emergence of order among coupled oscillators remains a cornerstone of nonlinear dynamics. At the quantum level, similar collective behavior often manifests in the spin degree of freedom, where the individual magnetic moments align to form macroscopic coherent states. In the presence of a transverse magnetic field, any spin-endowed particle undergoes Larmor precession -- a steady, periodic rotation around the field lines~\cite{spin_book2001}. Mastering the manipulation of the spin states is critical for the advancement of high-resolution imagining spectroscopy~\cite{emsley2013high}, spintronic devices~\cite{Yang_NatRevPhys2021}, and scalable quantum information processing~\cite{Xin_ChiPhysB2018}. All-optical control of spinor degrees of freedom has proven particularly effective, enabling the driving of spin precession in dilute atomic gases~\cite{opt_driv_PhysRevLett.6.280} and cold-atom Bose-Einstein condensates~\cite{PhysRevLett.95.050401_larmor_bose}, with direct implications for high-precision magnetometry~\cite{gas_magnetometer_nature}.

Microcavity polaritons offer a unique platform where the concepts of interconnected system synchronization and spin control converge. Exciton-polaritons (hereinafter polaritons) are bosonic quasiparticles emerging from the strong light-matter coupling between cavity photons and quantum-well excitons~\cite{kavokin_microcavities_2007}. Polariton condensates possess a spinor degree of freedom (corresponding to $\sigma^+$ and $\sigma^-$ circular polarization of the emitted light) and pronounced nonlinearities~\cite{ Kuriakose_NatPho2022}, enabling a rich landscape of spin phenomena including bistability~\cite{pickup_optical_2018, Ohadi19, Sigurdsson_PRR2020}, bifurcations~\cite{ohadi_spontaneous_2015} and the optical spin Hall effect~\cite{leyder_observation_2007}. Critically, the polariton condensate pseudo-spin can be manipulated optically via the excitation polarization, circumventing the need for integrated magnetic fields~\cite{Walker_PRL2011, PhysRevB.102.125419, baryshev_prl}.

Beyond single-site dynamics, repulsive outward polariton flow~\cite{PhysRevB.77.115340} and macroscopic spatial coherence allows coupling between condensates~\cite{PhysRevLett.100.170401,PhysRevX.6.031032,PhysRevLett.114.193901} which scales polariton systems into artificial lattices~\cite{Amo_2016,Alyatkin_Penrose} and coupled networks~\cite{Töpfer2020} for potential realization of high-speed analog simulators~\cite{PhysRevApplied.17.024063}. Under nonresonant excitation, the optically-induced repulsive potential from the exciton reservoir drives ballistic expansion of polaritons, enabling long-range coupling between the distant sites. Notably, optically trapped polaritons exhibit suppressed depolarization and decoherence in contrast to those residing directly atop the exciton reservoir. While the trapped condensates are spatially confined, the finite radial outflow of polaritons remains sufficient to induce mutual phase-locking~\cite{PhysRevB.101.155402}.  

Moreover, the presence of intrinsic spinor-dependent energy-splitting in microcavity polaritons has been shown to act as an effective magnetic field, inducing a self-induced Larmor precession of the optically trapped polariton condensate pseudo-spin under nonresonant, elliptically polarized excitation~\cite{baryshev_prl}. This pseudo-spin dynamics manifests as a periodic modulation of the polarization-filtered intensity correlation function. While the coherence time of optically trapped condensate is about $1\,\text{ns}$, the typical dephasing times of self-induced Larmor precessions reach up to $10\,\text{ns}$. Furthermore, by implementing an external optical stirring of the condensate at a frequency resonant with the precession, the stability of the pseudo-spin oscillations can be enhanced by a factor of nearly 20~\cite{gnusov2024observation}. 

In this Letter, we experimentally demonstrate the transfer of Larmor precession in a pair of distant, optically trapped polariton condensates one of which is externally stirred. We identify the amplitude of the intensity cross-correlation function as a metric for the inter-condensate synchronization. Our results reveal that the coupling, mediated by radial outflow, is critically dependent on both spectral overlap of the condensates photoluminescence and the mutual winding direction of the precessing pseudo-spins. Finally, we conclusively show that an optical stirring of a single ``master'' condensate resonantly drives the global synchronization of the condensate dimer, extending the pseudo-spin dephasing time of the coupled system by over an order of magnitude.

\begin{figure}[t]
\includegraphics [width=1\columnwidth]{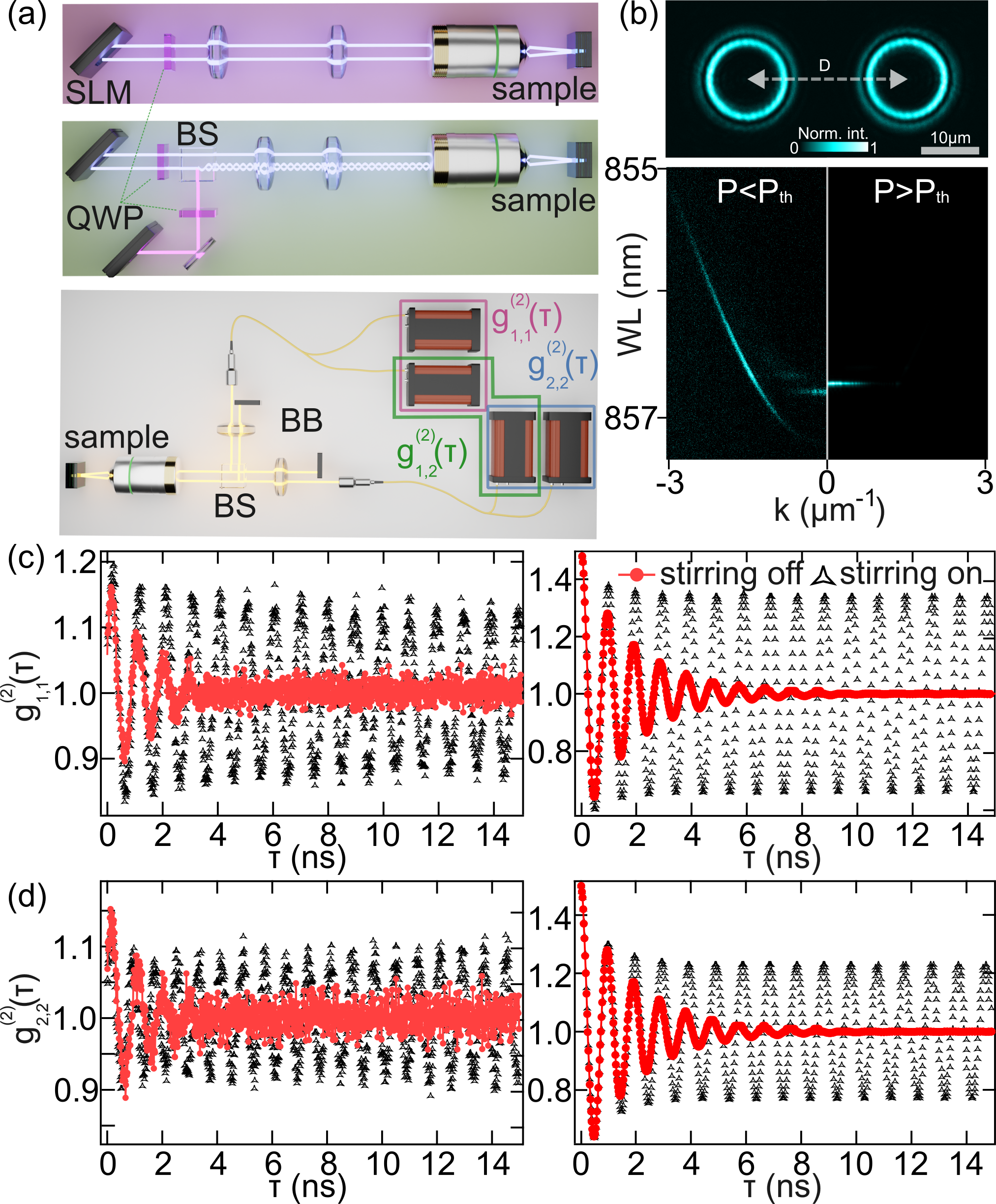}
\centering
\caption{(a) Experimental scheme for polariton condensate intensity correlation measurements. Two condensates ($C_1$, $C_2$) are excited in optical traps at a variable distance $D$ under two regimes: both in identical static traps (top) or with one trap stirred and the other static (middle). The bottom panel outlines the auto- and cross-correlation measurement setup, using color boxed detectors for corresponding correlation functions. BS - 50:50 beam splitter; BB - beam block. (b) Excitation profile (top panel). Typical dispersion image for the ground state trapped condensate below and above the threshold (bottom panel); WL - wavelength. (c)(d) Measured (left) and simulated (right) auto-correlation function $g^{(2)}_{1,1}(\tau)$ and $g^{(2)}_{2,2}(\tau)$ respectively for $C_1$ and $C_2$ in the static (red circles) and stirred (black triangles) traps.}
\label{fig.1}
\end{figure}

To probe the synchronization effects in an optically trapped condensate dimer we nonresonantly excite GaAs/AlAs$_{0.98}$P$_{0.02}$ 2$\lambda$-microcavity with embedded InGaAs quantum wells~\cite{doi:10.1063/1.4901814}. The sample is held at $4\,\text{K}$ within a cryostat, and the exciton-photon detuning is set to $-3\,\text{meV}$ in all experiments. The continuous-wave (CW) single-mode laser excitation is acousto-optically modulated using a square waveform at a $1\,\text{kHz}$ repetition rate and a $1\%$ duty cycle. Two independently programmable spatial light modulators (SLMs) are employed to control the excitation patterns for each condensate site. In the experiment, we realize two excitation regimes: (1) when both polariton condensates are confined within identical static traps, and (2) when one of the condensates in the dimer is stirred, while the second is in a static potential, as schematically illustrated in Fig.~\ref{fig.1}(a). The diameter of the static and stirred traps is set to $d\approx12~\mu m$ to excite polariton condensates in the trap ground state, while the distance between the traps $D$ can be varied, as depicted in Fig.~\ref{fig.1}(b). The optical stirring technique is based on the interference of two frequency-detuned CW lasers operating at $\omega_1$ and $\omega_2$. Using the SLMs we transform the laser beams into "perfect vortices"~\cite{chen2013dynamics} of opposite orbital angular momenta ($l_{1,2} = \pm 1$), whose coaxial superposition creates an elliptical potential rotating at angular frequency $\omega_{\text{rot}} = (\omega_1 - \omega_2) / 2$. Above the condensation threshold power $P_{\text{th}}$, polaritons localize inside the traps, populating the respective ground states. By modulating the phase masks on the SLMs, we precisely control the pump power and thus the condensates energies, while quarter-wave plates (QWPs) positioned after the SLMs allow for independent tuning of the excitation polarization for each site.

This independent polarization control is critical, as the polarization of the excitation beam dictates the initial polarization properties and subsequent pseudo-spin dynamics of each condensate~\cite{PhysRevB.102.125419}. Specifically, introduction of ellipticity to the pump polarization creates a spin imbalance in the exciton reservoir that is transferred directly to the condensate through the stimulated scattering process. This spin asymmetry generates an effective out-of-plane magnetic field component, whose magnitude and sign are precisely tunable via the pump ellipticity~\cite{PhysRevLett.89.077402, Shelykh_PRB2004, Krizhanovskii_PRB2006, Solnyshkov2007, Ryzhov_PhysRevRes2020, PhysRevB.99.165311}. The out-of-plane field does not act in isolation but coexists with a in-plane effective magnetic field arising from the microcavity's structural birefringence, which splits the energies of the orthogonal linear polarizations. The sum of these two fields results in a tilted total effective magnetic field. This field exerts a continuous torque on the condensate pseudo-spin driving the system into a dynamical regime of self-induced Larmor precession~\cite{Solnyshkov2007, PhysRevLett.129.155301, baryshev_prl}. As the pseudo-spin vector sweeps along its periodic trajectory, its projection onto the equatorial plane of the Poincaré sphere oscillates, manifesting experimentally as GHz-scale intensity modulations in the linear components of the emitted photoluminescence.  

In order to measure polariton condensate pseudo-spin precession (at the GHz frequency range) we utilize polarization resolved Hanbury-Brown-Twiss (HBT) interferometry. This technique enables direct measurement of the second-order intensity correlation functions,
\begin{align}\label{eq.2}
g_{i,j}^{(2)}(\tau)= \frac{\langle{a_{i}^{\dagger}(t)}{a_{j}^{\dagger}(t+\tau)}{a_{j}(t+\tau)}{a_{i}(t)}\rangle}{\langle{a_{i}^{\dagger}(t)}{a_{i}(t)}\rangle\langle{a_{j}^{\dagger}(t+\tau)}{a_{j}(t+\tau)}\rangle},
\end{align}
where $a^\dagger$ and $a$ are photon creation and annihilation operators for given condensate index $i,j\in1,2$. Thereby, $g_{1,1}^{(2)}(\tau)$ and $g_{2,2}^{(2)}(\tau)$ correspond to the polarization filtered (horizontal) intensity auto-correlation function for each condensate, while $g_{1,2}^{(2)}(\tau)$ denotes the cross-condensate correlation function. The decay envelop of such functions indicates how fast the condensate pseudo-spin precession dephases in time.

We create a pair of optically trapped condensates at an inter-site separation of $D = 21~\mu$m, under elliptically polarized excitation at $3P_{th}$. The excitation ellipticity and power are tuned such that the intrinsic pseudo-spin precession frequency for each site is matched. As illustrated in Fig.~\ref{fig.1}(c,d), the intensity auto-correlation functions $g_{i,i}^{(2)}(\tau)$ (i=1 or 2) for both trapped condensates created in the static traps (red curves) exhibit rapidly decaying oscillations, which happen due to the self-induced Larmor precession of the condensate pseudo-spin. In the measured dataset, these oscillations decay into noise after $4\text{ ns}$. While this decay time varies with polarization pinning conditions, it consistently remains under a few nanoseconds. When we turn on the optical stirring for condensate $C_1$ at the Larmor frequency (1 GHz for the shown data), the pseudo-spin dephasing time increases drastically compared to the static case, as evidenced by the persistent oscillations (black curve in Fig.~\ref{fig.1} (c)). Such prolongation is directly connected to the mutual frequency stability of two lasers employed to optically stir the $C_1$ condensate and can extend to $174 \pm 10$ ns~\cite{gnusov2024observation}. Remarkably, this enhancement is not spatially localized to the driven site. The auto-correlation function of the neighboring undriven condensate, $C_2$, exhibits an identical prolongation of pseudo-spin dephasing time whenever $C_1$ is stirred (black curve in Fig.~\ref{fig.1} (d)). This non-local stabilization indicates a robust "master-slave" entrainment, when the polariton outflow from the driven $C_1$ site acts as a coherent seed for $C_2$. 

This experimental observation is qualitatively reproduced by numerical simulations based on Adler equations for the mutual phases of the circular polarizations of the condensates. These equations are derived from a more general model for the slowly varying amplitudes of the polarizations of the first and the second condensates. The simulations are developed under the assumption that the polarization intensities are governed by energy balance and can only undergo small oscillations around their mean values (Kuramoto approximation). This approach was introduced in~\cite{lxv5-qwxw} to describe the dynamics of a single condensate driven by a rotating pump. In Sec. I of the Supplementary Materials (SM), we extend this approach to two interacting condensates and also show how the correlation functions can be expressed in terms of the mutual phases. In the SM, we further link the changes in the behavior of the correlation functions to switches between different synchronization regimes.

\begin{figure}[t]
\includegraphics [width=1\columnwidth]{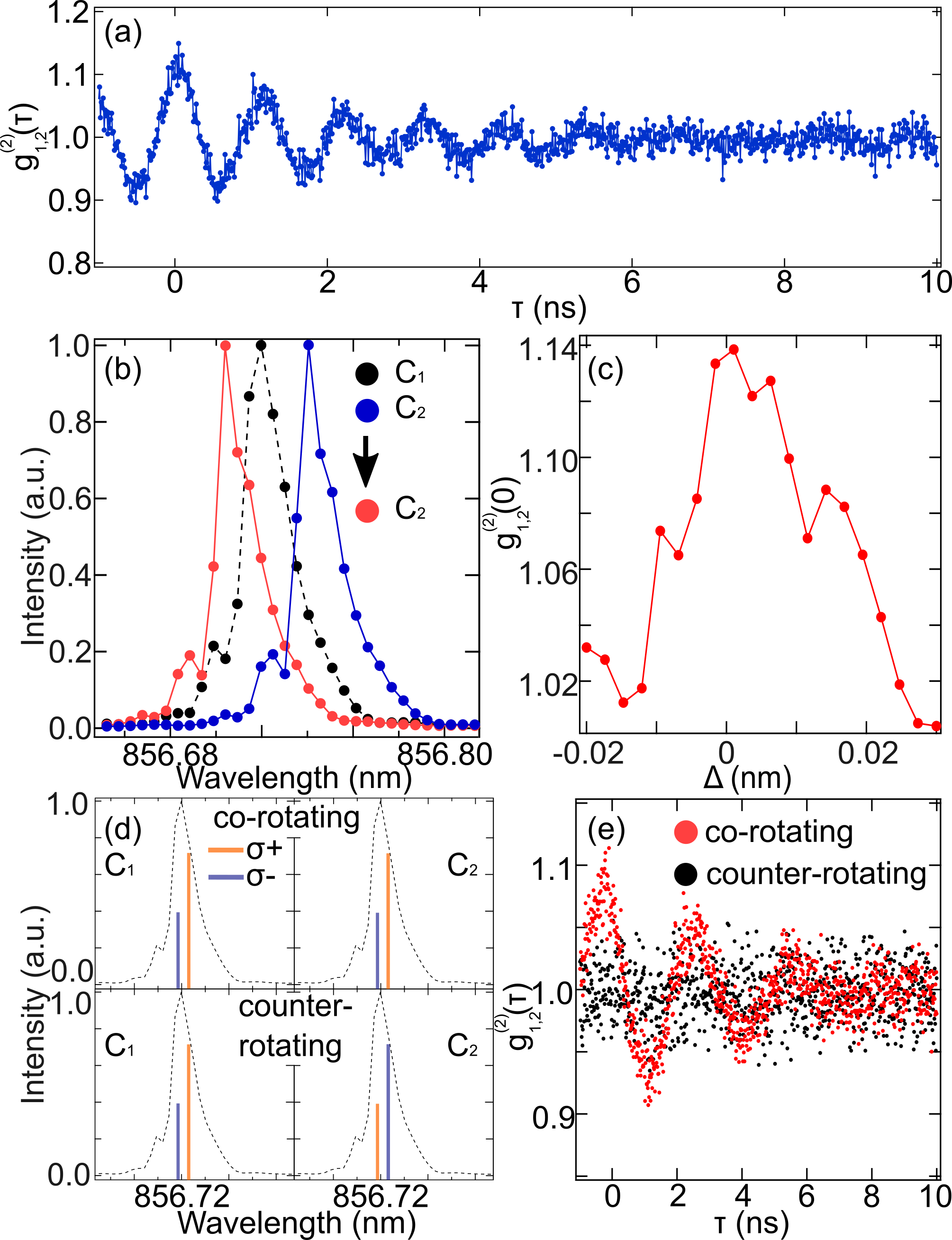}
\centering
\caption{(a) Measured polarization-filtered intensity cross-correlation  function $g^{(2)}_{1,2}(\tau)$ between two condensates. (b) Measured emission spectra of the of the condensate $C_1$ (black dots), while the emission wavelength of $C_2$ is tuned by varying the trap diameter from blue to red dots. (c) Zero-delay cross-correlation amplitude $g^{(2)}_{1,2}(0)$ as a function of the spectral detuning $\Delta$ between the two condensates. (d) Schematic representation of the co-circularly and counter-circularly polarized dimer spectra. The spectrum of the $C_1$ and $C_2$ are the same and constituent $\sigma+$ and $\sigma-$ lines are in resonance (upper panels) and out of resonance (lower panels). (e) Cross-correlation function $g^{(2)}_{1,2}(\tau)$ co-rotating (red) and counter-rotating (black) configurations of the condensate pseudo-spin precession.}
\label{fig.2} 
\end{figure}

To further investigate the interaction between the two condensates, we revert to the simpler case of two static traps. Analysis of the cross-condensate correlation function $g_{1,2}^{(2)}(\tau)$ reveals a clear signature of condensate coupling resulting in mutual synchronization between the two pseudo-spin precessions. As shown in Fig.~\ref{fig.2}(a), even in the absence of artificial optical stirring, the cross-correlation function exhibits periodic features, indicating that both condensate pseudo-spin projections are phase-locked. This synchronization, however, is significantly suppressed when condensates are spectrally detuned. To evaluate the intensity cross-correlation amplitude as a quantitative coupling metric, we tuned the emission wavelength detuning, $\Delta=\lambda_1-\lambda_2$, between condensates $C_2$ and $C_1$ (see Fig.~\ref{fig.2}(b)) by slightly varying the diameter of the $C_2$ confining potential. This tuning method was chosen over the conventional approach of varying excitation power, as the latter would shift the precession frequency by altering the induced spin imbalance. Importantly, the relationship between the potential size and emission wavelength for $C_2$ was calibrated in the absence of the neighboring condensate to eliminate coupling-induced energy shifts. As shown in Fig.~\ref{fig.2}(c), the measured zero-delay cross-correlation $g^{(2)}_{1,2}(0)$ reaches its maximum when the spectral detuning $\Delta$ is minimized. However, the precession synchronization is rapidly lost as the system is tuned away from this resonance condition. Furthermore, we investigate the role of precession "winding" by rotating the QWP to switch the excitation polarization of $C_2$ from clockwise to counter-clockwise while maintaining a clockwise polarization of the excitation for $C_1$. In this configuration, both the integrated emission spectrum and Larmor frequency of $C_2$ match those of $C_1$. However, the sign of the spin-imbalance-induced energy splitting is opposite (see Fig.~\ref{fig.2}(d)), which effectively drives the precession in the opposite direction. Crucially, we observe that when the pseudo-spins precess with opposite winding, the coupling is entirely suppressed, as evidenced by the lack of correlation in Fig.~\ref{fig.2}(e).

\begin{figure}[t]
\includegraphics [width=1\columnwidth]{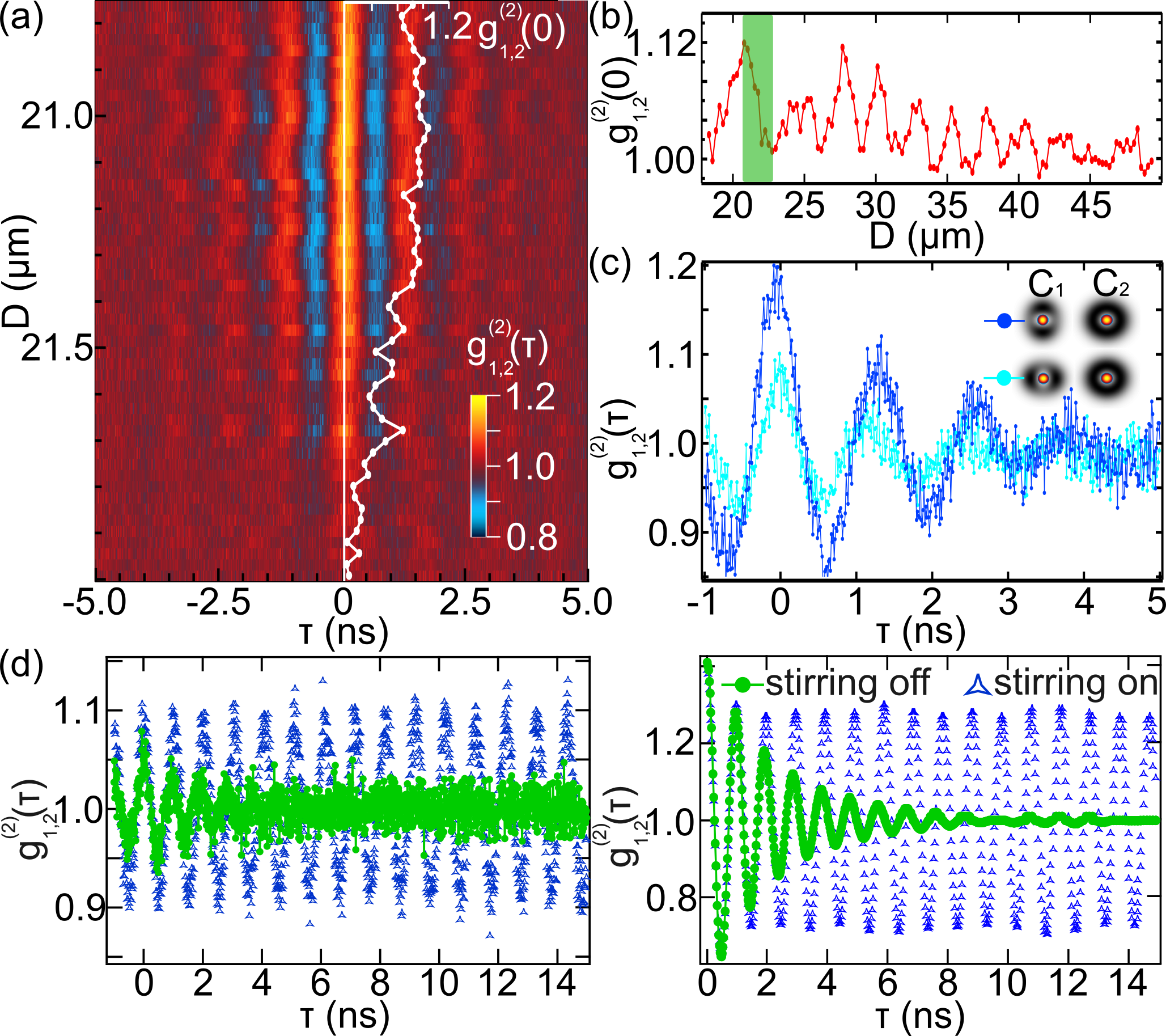}
\centering
\caption{(a) The measured map of the cross-correlation function $g^{(2)}_{1,2}(\tau)$ as a function of $D$. White dots show the correlation amplitude at zero-delay, $g^{(2)}_{1,2}(0)$ as a function of $D$. (b) Zero-delay cross-correlation amplitude versus the extended range of $D$. Highlighted area corresponds to the range presented in panel (a). (c) Intensity cross-correlation functions $g^{(2)}_{1,2}(\tau)$ for trap configurations schematically shown in the inset. (e) Measured (left) and simulated (right) cross-correlation functions $g^{(2)}_{1,2}(\tau)$ when the optical stirring of $C_1$ is off (green circles) and on (blue triangles). The simulations are done withing the framework of the Kuramoto approximation introduced in the SM.}

\label{fig.3}
\end{figure}

The observed sensitivity to both energy matching and direction of the precession winding highlights that synchronization is not merely a consequence of spatial proximity, but a resonant exchange of spin-polarized particles. The propagation-dependent phase implies that the synchronization strength must vary with the physical separation of the traps~\cite{dovzhenko2023next}. Indeed, Fig.~\ref{fig.3}(a) shows that as the spatial separation is increased, the zero-delay cross-condensate correlation amplitude $g_{1,2}^{(2)}(0)$ monotonically decreases, reaching a point of total asynchronicity at $D = 21.8~\mu$m. Furthermore, measuring $g_{1,2}^{(2)}(0)$ over a wider spatial range evidences periodic revivals of the synchronization, as shown in Fig.~\ref{fig.3}(b). These revivals point towards the existing phase-locking mechanism governed by the constructive and destructive interference of the bidirectional polariton outflow between the two condensates. 

The implementation of an optically imprinted excitation potential provides a versatile platform to further manipulate condensate coupling through dynamic reconfiguration of the trap landscape. By introducing in-plane axial asymmetries, specifically through the elongation of the trap along a preferred axis, we break the rotational symmetry of the system. This geometric anisotropy, coupled with the intrinsic transverse electric-transverse magnetic (TE-TM) splitting of the microcavity, lifts the energy degeneracy of the trap's ground state, favoring a distinct pseudo-spin orientation~\cite{gnusov_prapl}. Consequently, the steady-state condensate exhibits a high degree of linear polarization aligned with the short axis of the elliptical trap, physically analogous to linear dichroism. From a dynamical perspective, the spatial confinement is weakest along the short axis of the potential. This reduction in the effective potential barrier facilitates an enhanced lateral flow of polaritons along this direction. In a dimer configuration, this directional flow significantly strengthens the coupling between adjacent condensates. The increased overlap of the respective wavefunctions enhances phase-locking and inter-trap correlations, manifested as a higher value of the zero-delay cross-condensate correlation function $g_{1,2}^{(2)}(0)$ as shown in Fig.~\ref{fig.3}(c). We compare cases where polariton flow is either geometrically enhanced or suppressed. The coupling strength is higher when the traps are oriented to favor mutual flow (dark blue dots in Fig.~\ref{fig.3}(c)). In contrast, when the traps are aligned such that the higher potential barrier intervenes, the coupling is suppressed (light blue line in Fig.~\ref{fig.3}(c)). These results demonstrate that the geometrical tuning of the trap not only selects the polarization state but also serves as a control knob for the synchronization between localized polariton condensates.
\par The optical stirring rotates such symmetry-broken optical trap averaging the amplitude of the cross-correlation function to the intermediate value. The extension of the pseudo-spin dephasing time in $C_2$ during the optical stirring of $C_1$ (Fig.~\ref{fig.1}(c)) indicates a robust, unidirectional transfer of stability. In this configuration, the driven condensate $C_1$ acts as a ``master'' oscillator, whose outflow imposes a coherent periodic modulation on $C_2$. This external modulation effectively seeds the Larmor precession of the ``slave'' condensate site, entraining its pseudo-spin frequency to that of the ``master'' drive. 

Beyond simple frequency matching, cross-condensate correlation measurements prove that the outflow carries the phase-locked timing of the master condensate. We confirm global phase-locking with $g^{(2)}_{1,2}(\tau)$ measurement and simulation based on the coupled mode approach~\cite{gnusov2024observation}, which reveals a significant prolongation of mutual synchronization (see Fig.~\ref{fig.3}(e)). Additionally, we make sure that the observed effect is not a cross-talk of condensates in our detection scheme (see SM Sec.II). Our model further expands the range of parameters where the synchronization is expected and explore how correlation function evolves when measured in different polarization basis (see SM for details).  Crucially, in this ``master-slave'' regime, the system overcomes the intrinsic temporal limits of self-induced precession. The stability of the pseudo-spin synchronization becomes decoupled from the stochastic fluctuations inherent to the polariton lifetime and first-order phase coherence and instead, it is governed by the superior frequency stability of the driving excitation lasers. These results demonstrate that a single clock condensate can support synchronization of an entire polaritonic network, providing a scalable architecture for high-speed spinoptronic signal distribution.

In conclusion, we have explored the synchronization of self-induced Larmor precessions in a dimer of optically trapped polariton condensates. Our findings reveal that the coupling strength can be quantified by the cross-correlation amplitude $g^{(2)}_{1,2}(0)$. We demonstrated that this synchronization is a resonant process requiring both energy matching and the same precession winding direction. By introducing in-plane trap asymmetry and dynamic optical stirring to one of the sites, we examined how geometric and temporal control of the confining potential can be used to engineer and drastically extend the pseudo-spin dephasing time of coupled condensates. These results have significant implications for the development of polaritonic circuits. The ability of a driven ``master'' condensate to impose its pseudo-spin precession phase onto an undriven ``slave'' site provides a route for long-range, all-optical distribution of clock signals in spinoptronic architectures. Looking forward, the scalability of this platform to larger 1D and 2D lattices opens the door to simulating complex non-equilibrium synchronization phenomena in a solid-state environment. 

This work was supported by Russian Science Foundation (RSF) grant no. 24-72-10118, https://rscf.ru/en/project/24-72-10118/.


\begin{thebibliography}{40}%
\makeatletter
\providecommand \@ifxundefined [1]{%
 \@ifx{#1\undefined}
}%
\providecommand \@ifnum [1]{%
 \ifnum #1\expandafter \@firstoftwo
 \else \expandafter \@secondoftwo
 \fi
}%
\providecommand \@ifx [1]{%
 \ifx #1\expandafter \@firstoftwo
 \else \expandafter \@secondoftwo
 \fi
}%
\providecommand \natexlab [1]{#1}%
\providecommand \enquote  [1]{``#1''}%
\providecommand \bibnamefont  [1]{#1}%
\providecommand \bibfnamefont [1]{#1}%
\providecommand \citenamefont [1]{#1}%
\providecommand \href@noop [0]{\@secondoftwo}%
\providecommand \href [0]{\begingroup \@sanitize@url \@href}%
\providecommand \@href[1]{\@@startlink{#1}\@@href}%
\providecommand \@@href[1]{\endgroup#1\@@endlink}%
\providecommand \@sanitize@url [0]{\catcode `\\12\catcode `\$12\catcode `\&12\catcode `\#12\catcode `\^12\catcode `\_12\catcode `\%12\relax}%
\providecommand \@@startlink[1]{}%
\providecommand \@@endlink[0]{}%
\providecommand \url  [0]{\begingroup\@sanitize@url \@url }%
\providecommand \@url [1]{\endgroup\@href {#1}{\urlprefix }}%
\providecommand \urlprefix  [0]{URL }%
\providecommand \Eprint [0]{\href }%
\providecommand \doibase [0]{http://dx.doi.org/}%
\providecommand \selectlanguage [0]{\@gobble}%
\providecommand \bibinfo  [0]{\@secondoftwo}%
\providecommand \bibfield  [0]{\@secondoftwo}%
\providecommand \translation [1]{[#1]}%
\providecommand \BibitemOpen [0]{}%
\providecommand \bibitemStop [0]{}%
\providecommand \bibitemNoStop [0]{.\EOS\space}%
\providecommand \EOS [0]{\spacefactor3000\relax}%
\providecommand \BibitemShut  [1]{\csname bibitem#1\endcsname}%
\let\auto@bib@innerbib\@empty
%</preamble>
\bibitem [{\citenamefont {Pikovsky}\ \emph {et~al.}(2001)\citenamefont {Pikovsky}, \citenamefont {Rosenblum},\ and\ \citenamefont {Kurths}}]{Pikovsky_Rosenblum_Kurths_2001}%
  \BibitemOpen
  \bibfield  {author} {\bibinfo {author} {\bibfnamefont {A.}~\bibnamefont {Pikovsky}}, \bibinfo {author} {\bibfnamefont {M.}~\bibnamefont {Rosenblum}}, \ and\ \bibinfo {author} {\bibfnamefont {J.}~\bibnamefont {Kurths}},\ }\href@noop {} {\emph {\bibinfo {title} {Synchronization: A Universal Concept in Nonlinear Sciences}}},\ Cambridge Nonlinear Science Series\ (\bibinfo  {publisher} {Cambridge University Press},\ \bibinfo {year} {2001})\BibitemShut {NoStop}%
\bibitem [{\citenamefont {Levitt}(2013)}]{spin_book2001}%
  \BibitemOpen
  \bibfield  {author} {\bibinfo {author} {\bibfnamefont {M.~H.}\ \bibnamefont {Levitt}},\ }\href@noop {} {\emph {\bibinfo {title} {Spin dynamics: basics of nuclear magnetic resonance}}}\ (\bibinfo  {publisher} {John Wiley \& Sons},\ \bibinfo {address} {Chichester, UK},\ \bibinfo {year} {2013})\BibitemShut {NoStop}%
\bibitem [{\citenamefont {Emsley}\ \emph {et~al.}(1966)\citenamefont {Emsley}, \citenamefont {Feeney},\ and\ \citenamefont {Sutcliffe}}]{emsley2013high}%
  \BibitemOpen
  \bibfield  {author} {\bibinfo {author} {\bibfnamefont {J.~W.}\ \bibnamefont {Emsley}}, \bibinfo {author} {\bibfnamefont {J.}~\bibnamefont {Feeney}}, \ and\ \bibinfo {author} {\bibfnamefont {L.~H.}\ \bibnamefont {Sutcliffe}},\ }\href@noop {} {\emph {\bibinfo {title} {High Resolution Nuclear Magnetic Resonance Spectroscopy}}}\ (\bibinfo  {publisher} {Pergamon Press},\ \bibinfo {address} {Oxford, UK},\ \bibinfo {year} {1966})\BibitemShut {NoStop}%
\bibitem [{\citenamefont {Yang}\ \emph {et~al.}(2021)\citenamefont {Yang}, \citenamefont {Naaman}, \citenamefont {Paltiel},\ and\ \citenamefont {Parkin}}]{Yang_NatRevPhys2021}%
  \BibitemOpen
  \bibfield  {author} {\bibinfo {author} {\bibfnamefont {S.-H.}\ \bibnamefont {Yang}}, \bibinfo {author} {\bibfnamefont {R.}~\bibnamefont {Naaman}}, \bibinfo {author} {\bibfnamefont {Y.}~\bibnamefont {Paltiel}}, \ and\ \bibinfo {author} {\bibfnamefont {S.~S.~P.}\ \bibnamefont {Parkin}},\ }\href {\doibase 10.1038/s42254-021-00302-9} {\bibfield  {journal} {\bibinfo  {journal} {Nature Reviews Physics}\ }\textbf {\bibinfo {volume} {3}},\ \bibinfo {pages} {328} (\bibinfo {year} {2021})}\BibitemShut {NoStop}%
\bibitem [{\citenamefont {Xin}\ \emph {et~al.}(2018)\citenamefont {Xin}, \citenamefont {Wang}, \citenamefont {Li}, \citenamefont {Kong}, \citenamefont {Wei}, \citenamefont {Wang}, \citenamefont {Ruan},\ and\ \citenamefont {Long}}]{Xin_ChiPhysB2018}%
  \BibitemOpen
  \bibfield  {author} {\bibinfo {author} {\bibfnamefont {T.}~\bibnamefont {Xin}}, \bibinfo {author} {\bibfnamefont {B.-X.}\ \bibnamefont {Wang}}, \bibinfo {author} {\bibfnamefont {K.-R.}\ \bibnamefont {Li}}, \bibinfo {author} {\bibfnamefont {X.-Y.}\ \bibnamefont {Kong}}, \bibinfo {author} {\bibfnamefont {S.-J.}\ \bibnamefont {Wei}}, \bibinfo {author} {\bibfnamefont {T.}~\bibnamefont {Wang}}, \bibinfo {author} {\bibfnamefont {D.}~\bibnamefont {Ruan}}, \ and\ \bibinfo {author} {\bibfnamefont {G.-L.}\ \bibnamefont {Long}},\ }\href {\doibase 10.1088/1674-1056/27/2/020308} {\bibfield  {journal} {\bibinfo  {journal} {Chinese Physics B}\ }\textbf {\bibinfo {volume} {27}},\ \bibinfo {pages} {020308} (\bibinfo {year} {2018})}\BibitemShut {NoStop}%
\bibitem [{\citenamefont {Bell}\ and\ \citenamefont {Bloom}(1961)}]{opt_driv_PhysRevLett.6.280}%
  \BibitemOpen
  \bibfield  {author} {\bibinfo {author} {\bibfnamefont {W.~E.}\ \bibnamefont {Bell}}\ and\ \bibinfo {author} {\bibfnamefont {A.~L.}\ \bibnamefont {Bloom}},\ }\href {\doibase 10.1103/PhysRevLett.6.280} {\bibfield  {journal} {\bibinfo  {journal} {Phys. Rev. Lett.}\ }\textbf {\bibinfo {volume} {6}},\ \bibinfo {pages} {280} (\bibinfo {year} {1961})}\BibitemShut {NoStop}%
\bibitem [{\citenamefont {Higbie}\ \emph {et~al.}(2005)\citenamefont {Higbie}, \citenamefont {Sadler}, \citenamefont {Inouye}, \citenamefont {Chikkatur}, \citenamefont {Leslie}, \citenamefont {Moore}, \citenamefont {Savalli},\ and\ \citenamefont {Stamper-Kurn}}]{PhysRevLett.95.050401_larmor_bose}%
  \BibitemOpen
  \bibfield  {author} {\bibinfo {author} {\bibfnamefont {J.~M.}\ \bibnamefont {Higbie}}, \bibinfo {author} {\bibfnamefont {L.~E.}\ \bibnamefont {Sadler}}, \bibinfo {author} {\bibfnamefont {S.}~\bibnamefont {Inouye}}, \bibinfo {author} {\bibfnamefont {A.~P.}\ \bibnamefont {Chikkatur}}, \bibinfo {author} {\bibfnamefont {S.~R.}\ \bibnamefont {Leslie}}, \bibinfo {author} {\bibfnamefont {K.~L.}\ \bibnamefont {Moore}}, \bibinfo {author} {\bibfnamefont {V.}~\bibnamefont {Savalli}}, \ and\ \bibinfo {author} {\bibfnamefont {D.~M.}\ \bibnamefont {Stamper-Kurn}},\ }\href {\doibase 10.1103/PhysRevLett.95.050401} {\bibfield  {journal} {\bibinfo  {journal} {Phys. Rev. Lett.}\ }\textbf {\bibinfo {volume} {95}},\ \bibinfo {pages} {050401} (\bibinfo {year} {2005})}\BibitemShut {NoStop}%
\bibitem [{\citenamefont {Kominis}\ \emph {et~al.}(2003)\citenamefont {Kominis}, \citenamefont {Kornack}, \citenamefont {Allred},\ and\ \citenamefont {Romalis}}]{gas_magnetometer_nature}%
  \BibitemOpen
  \bibfield  {author} {\bibinfo {author} {\bibfnamefont {I.~K.}\ \bibnamefont {Kominis}}, \bibinfo {author} {\bibfnamefont {T.~W.}\ \bibnamefont {Kornack}}, \bibinfo {author} {\bibfnamefont {J.~C.}\ \bibnamefont {Allred}}, \ and\ \bibinfo {author} {\bibfnamefont {M.~V.}\ \bibnamefont {Romalis}},\ }\href {\doibase 10.1038/nature01484} {\bibfield  {journal} {\bibinfo  {journal} {Nature}\ }\textbf {\bibinfo {volume} {422}},\ \bibinfo {pages} {596} (\bibinfo {year} {2003})}\BibitemShut {NoStop}%
\bibitem [{\citenamefont {Kavokin}\ \emph {et~al.}(2007)\citenamefont {Kavokin}, \citenamefont {Baumberg}, \citenamefont {Malpuech},\ and\ \citenamefont {Laussy}}]{kavokin_microcavities_2007}%
  \BibitemOpen
  \bibfield  {author} {\bibinfo {author} {\bibfnamefont {A.}~\bibnamefont {Kavokin}}, \bibinfo {author} {\bibfnamefont {J.~J.}\ \bibnamefont {Baumberg}}, \bibinfo {author} {\bibfnamefont {G.}~\bibnamefont {Malpuech}}, \ and\ \bibinfo {author} {\bibfnamefont {F.~P.}\ \bibnamefont {Laussy}},\ }\href@noop {} {\emph {\bibinfo {title} {Microcavities}}}\ (\bibinfo  {publisher} {Oxford University Press},\ \bibinfo {address} {Oxford},\ \bibinfo {year} {2007})\BibitemShut {NoStop}%
\bibitem [{\citenamefont {Kuriakose}\ \emph {et~al.}(2022)\citenamefont {Kuriakose}, \citenamefont {Walker}, \citenamefont {Dowling}, \citenamefont {Kyriienko}, \citenamefont {Shelykh}, \citenamefont {St-Jean}, \citenamefont {Zambon}, \citenamefont {Lema{\^i}tre}, \citenamefont {Sagnes}, \citenamefont {Legratiet}, \citenamefont {Harouri}, \citenamefont {Ravets}, \citenamefont {Skolnick}, \citenamefont {Amo}, \citenamefont {Bloch},\ and\ \citenamefont {Krizhanovskii}}]{Kuriakose_NatPho2022}%
  \BibitemOpen
  \bibfield  {author} {\bibinfo {author} {\bibfnamefont {T.}~\bibnamefont {Kuriakose}}, \bibinfo {author} {\bibfnamefont {P.~M.}\ \bibnamefont {Walker}}, \bibinfo {author} {\bibfnamefont {T.}~\bibnamefont {Dowling}}, \bibinfo {author} {\bibfnamefont {O.}~\bibnamefont {Kyriienko}}, \bibinfo {author} {\bibfnamefont {I.~A.}\ \bibnamefont {Shelykh}}, \bibinfo {author} {\bibfnamefont {P.}~\bibnamefont {St-Jean}}, \bibinfo {author} {\bibfnamefont {N.~C.}\ \bibnamefont {Zambon}}, \bibinfo {author} {\bibfnamefont {A.}~\bibnamefont {Lema{\^i}tre}}, \bibinfo {author} {\bibfnamefont {I.}~\bibnamefont {Sagnes}}, \bibinfo {author} {\bibfnamefont {L.}~\bibnamefont {Legratiet}}, \bibinfo {author} {\bibfnamefont {A.}~\bibnamefont {Harouri}}, \bibinfo {author} {\bibfnamefont {S.}~\bibnamefont {Ravets}}, \bibinfo {author} {\bibfnamefont {M.~S.}\ \bibnamefont {Skolnick}}, \bibinfo {author} {\bibfnamefont {A.}~\bibnamefont {Amo}}, \bibinfo {author} {\bibfnamefont {J.}~\bibnamefont {Bloch}}, \ and\ \bibinfo {author}
  {\bibfnamefont {D.~N.}\ \bibnamefont {Krizhanovskii}},\ }\href {\doibase 10.1038/s41566-022-01019-6} {\bibfield  {journal} {\bibinfo  {journal} {Nature Photonics}\ }\textbf {\bibinfo {volume} {16}},\ \bibinfo {pages} {566} (\bibinfo {year} {2022})}\BibitemShut {NoStop}%
\bibitem [{\citenamefont {Pickup}\ \emph {et~al.}(2018)\citenamefont {Pickup}, \citenamefont {Kalinin}, \citenamefont {Askitopoulos}, \citenamefont {Hatzopoulos}, \citenamefont {Savvidis}, \citenamefont {Berloff},\ and\ \citenamefont {Lagoudakis}}]{pickup_optical_2018}%
  \BibitemOpen
  \bibfield  {author} {\bibinfo {author} {\bibfnamefont {L.}~\bibnamefont {Pickup}}, \bibinfo {author} {\bibfnamefont {K.}~\bibnamefont {Kalinin}}, \bibinfo {author} {\bibfnamefont {A.}~\bibnamefont {Askitopoulos}}, \bibinfo {author} {\bibfnamefont {Z.}~\bibnamefont {Hatzopoulos}}, \bibinfo {author} {\bibfnamefont {P.}~\bibnamefont {Savvidis}}, \bibinfo {author} {\bibfnamefont {N.}~\bibnamefont {Berloff}}, \ and\ \bibinfo {author} {\bibfnamefont {P.}~\bibnamefont {Lagoudakis}},\ }\href {\doibase 10.1103/PhysRevLett.120.225301} {\bibfield  {journal} {\bibinfo  {journal} {Physical Review Letters}\ }\textbf {\bibinfo {volume} {120}},\ \bibinfo {pages} {225301} (\bibinfo {year} {2018})}\BibitemShut {NoStop}%
\bibitem [{\citenamefont {del Valle-Inclan~Redondo}\ \emph {et~al.}(2019{\natexlab{a}})\citenamefont {del Valle-Inclan~Redondo}, \citenamefont {Sigurdsson}, \citenamefont {Ohadi}, \citenamefont {Shelykh}, \citenamefont {Rubo}, \citenamefont {Hatzopoulos}, \citenamefont {Savvidis},\ and\ \citenamefont {Baumberg}}]{Ohadi19}%
  \BibitemOpen
  \bibfield  {author} {\bibinfo {author} {\bibfnamefont {Y.}~\bibnamefont {del Valle-Inclan~Redondo}}, \bibinfo {author} {\bibfnamefont {H.}~\bibnamefont {Sigurdsson}}, \bibinfo {author} {\bibfnamefont {H.}~\bibnamefont {Ohadi}}, \bibinfo {author} {\bibfnamefont {I.~A.}\ \bibnamefont {Shelykh}}, \bibinfo {author} {\bibfnamefont {Y.~G.}\ \bibnamefont {Rubo}}, \bibinfo {author} {\bibfnamefont {Z.}~\bibnamefont {Hatzopoulos}}, \bibinfo {author} {\bibfnamefont {P.}~\bibnamefont {Savvidis}}, \ and\ \bibinfo {author} {\bibfnamefont {J.}~\bibnamefont {Baumberg}},\ }\href {\doibase 10.1103/PhysRevB.99.165311} {\bibfield  {journal} {\bibinfo  {journal} {Phys. Rev. B}\ }\textbf {\bibinfo {volume} {99}},\ \bibinfo {pages} {165311} (\bibinfo {year} {2019}{\natexlab{a}})}\BibitemShut {NoStop}%
\bibitem [{\citenamefont {Sigurdsson}(2020)}]{Sigurdsson_PRR2020}%
  \BibitemOpen
  \bibfield  {author} {\bibinfo {author} {\bibfnamefont {H.}~\bibnamefont {Sigurdsson}},\ }\href {\doibase 10.1103/PhysRevResearch.2.023323} {\bibfield  {journal} {\bibinfo  {journal} {Phys. Rev. Res.}\ }\textbf {\bibinfo {volume} {2}},\ \bibinfo {pages} {023323} (\bibinfo {year} {2020})}\BibitemShut {NoStop}%
\bibitem [{\citenamefont {Ohadi}\ \emph {et~al.}(2015)\citenamefont {Ohadi}, \citenamefont {Dreismann}, \citenamefont {Rubo}, \citenamefont {Pinsker}, \citenamefont {del Valle-Inclan~Redondo}, \citenamefont {Tsintzos}, \citenamefont {Hatzopoulos}, \citenamefont {Savvidis},\ and\ \citenamefont {Baumberg}}]{ohadi_spontaneous_2015}%
  \BibitemOpen
  \bibfield  {author} {\bibinfo {author} {\bibfnamefont {H.}~\bibnamefont {Ohadi}}, \bibinfo {author} {\bibfnamefont {A.}~\bibnamefont {Dreismann}}, \bibinfo {author} {\bibfnamefont {Y.}~\bibnamefont {Rubo}}, \bibinfo {author} {\bibfnamefont {F.}~\bibnamefont {Pinsker}}, \bibinfo {author} {\bibfnamefont {Y.}~\bibnamefont {del Valle-Inclan~Redondo}}, \bibinfo {author} {\bibfnamefont {S.}~\bibnamefont {Tsintzos}}, \bibinfo {author} {\bibfnamefont {Z.}~\bibnamefont {Hatzopoulos}}, \bibinfo {author} {\bibfnamefont {P.}~\bibnamefont {Savvidis}}, \ and\ \bibinfo {author} {\bibfnamefont {J.}~\bibnamefont {Baumberg}},\ }\href {\doibase 10.1103/PhysRevX.5.031002} {\bibfield  {journal} {\bibinfo  {journal} {Physical Review X}\ }\textbf {\bibinfo {volume} {5}},\ \bibinfo {pages} {031002} (\bibinfo {year} {2015})}\BibitemShut {NoStop}%
\bibitem [{\citenamefont {Leyder}\ \emph {et~al.}(2007)\citenamefont {Leyder}, \citenamefont {Romanelli}, \citenamefont {Karr}, \citenamefont {Giacobino}, \citenamefont {Liew}, \citenamefont {Glazov}, \citenamefont {Kavokin}, \citenamefont {Malpuech},\ and\ \citenamefont {Bramati}}]{leyder_observation_2007}%
  \BibitemOpen
  \bibfield  {author} {\bibinfo {author} {\bibfnamefont {C.}~\bibnamefont {Leyder}}, \bibinfo {author} {\bibfnamefont {M.}~\bibnamefont {Romanelli}}, \bibinfo {author} {\bibfnamefont {J.~P.}\ \bibnamefont {Karr}}, \bibinfo {author} {\bibfnamefont {E.}~\bibnamefont {Giacobino}}, \bibinfo {author} {\bibfnamefont {T.~C.~H.}\ \bibnamefont {Liew}}, \bibinfo {author} {\bibfnamefont {M.~M.}\ \bibnamefont {Glazov}}, \bibinfo {author} {\bibfnamefont {A.~V.}\ \bibnamefont {Kavokin}}, \bibinfo {author} {\bibfnamefont {G.}~\bibnamefont {Malpuech}}, \ and\ \bibinfo {author} {\bibfnamefont {A.}~\bibnamefont {Bramati}},\ }\href {\doibase 10.1038/nphys676} {\bibfield  {journal} {\bibinfo  {journal} {Nature Physics}\ }\textbf {\bibinfo {volume} {3}},\ \bibinfo {pages} {628} (\bibinfo {year} {2007})}\BibitemShut {NoStop}%
\bibitem [{\citenamefont {Walker}\ \emph {et~al.}(2011)\citenamefont {Walker}, \citenamefont {Liew}, \citenamefont {Sarkar}, \citenamefont {Durska}, \citenamefont {Love}, \citenamefont {Skolnick}, \citenamefont {Roberts}, \citenamefont {Shelykh}, \citenamefont {Kavokin},\ and\ \citenamefont {Krizhanovskii}}]{Walker_PRL2011}%
  \BibitemOpen
  \bibfield  {author} {\bibinfo {author} {\bibfnamefont {P.}~\bibnamefont {Walker}}, \bibinfo {author} {\bibfnamefont {T.~C.~H.}\ \bibnamefont {Liew}}, \bibinfo {author} {\bibfnamefont {D.}~\bibnamefont {Sarkar}}, \bibinfo {author} {\bibfnamefont {M.}~\bibnamefont {Durska}}, \bibinfo {author} {\bibfnamefont {A.~P.~D.}\ \bibnamefont {Love}}, \bibinfo {author} {\bibfnamefont {M.~S.}\ \bibnamefont {Skolnick}}, \bibinfo {author} {\bibfnamefont {J.~S.}\ \bibnamefont {Roberts}}, \bibinfo {author} {\bibfnamefont {I.~A.}\ \bibnamefont {Shelykh}}, \bibinfo {author} {\bibfnamefont {A.~V.}\ \bibnamefont {Kavokin}}, \ and\ \bibinfo {author} {\bibfnamefont {D.~N.}\ \bibnamefont {Krizhanovskii}},\ }\href {\doibase 10.1103/PhysRevLett.106.257401} {\bibfield  {journal} {\bibinfo  {journal} {Phys. Rev. Lett.}\ }\textbf {\bibinfo {volume} {106}},\ \bibinfo {pages} {257401} (\bibinfo {year} {2011})}\BibitemShut {NoStop}%
\bibitem [{\citenamefont {Gnusov}\ \emph {et~al.}(2020)\citenamefont {Gnusov}, \citenamefont {Sigurdsson}, \citenamefont {Baryshev}, \citenamefont {Ermatov}, \citenamefont {Askitopoulos},\ and\ \citenamefont {Lagoudakis}}]{PhysRevB.102.125419}%
  \BibitemOpen
  \bibfield  {author} {\bibinfo {author} {\bibfnamefont {I.}~\bibnamefont {Gnusov}}, \bibinfo {author} {\bibfnamefont {H.}~\bibnamefont {Sigurdsson}}, \bibinfo {author} {\bibfnamefont {S.}~\bibnamefont {Baryshev}}, \bibinfo {author} {\bibfnamefont {T.}~\bibnamefont {Ermatov}}, \bibinfo {author} {\bibfnamefont {A.}~\bibnamefont {Askitopoulos}}, \ and\ \bibinfo {author} {\bibfnamefont {P.~G.}\ \bibnamefont {Lagoudakis}},\ }\href {\doibase 10.1103/PhysRevB.102.125419} {\bibfield  {journal} {\bibinfo  {journal} {Phys. Rev. B}\ }\textbf {\bibinfo {volume} {102}},\ \bibinfo {pages} {125419} (\bibinfo {year} {2020})}\BibitemShut {NoStop}%
\bibitem [{\citenamefont {Baryshev}\ \emph {et~al.}(2022)\citenamefont {Baryshev}, \citenamefont {Zasedatelev}, \citenamefont {Sigurdsson}, \citenamefont {Gnusov}, \citenamefont {T\"opfer}, \citenamefont {Askitopoulos},\ and\ \citenamefont {Lagoudakis}}]{baryshev_prl}%
  \BibitemOpen
  \bibfield  {author} {\bibinfo {author} {\bibfnamefont {S.}~\bibnamefont {Baryshev}}, \bibinfo {author} {\bibfnamefont {A.}~\bibnamefont {Zasedatelev}}, \bibinfo {author} {\bibfnamefont {H.}~\bibnamefont {Sigurdsson}}, \bibinfo {author} {\bibfnamefont {I.}~\bibnamefont {Gnusov}}, \bibinfo {author} {\bibfnamefont {J.~D.}\ \bibnamefont {T\"opfer}}, \bibinfo {author} {\bibfnamefont {A.}~\bibnamefont {Askitopoulos}}, \ and\ \bibinfo {author} {\bibfnamefont {P.~G.}\ \bibnamefont {Lagoudakis}},\ }\href {\doibase 10.1103/PhysRevLett.128.087402} {\bibfield  {journal} {\bibinfo  {journal} {Phys. Rev. Lett.}\ }\textbf {\bibinfo {volume} {128}},\ \bibinfo {pages} {087402} (\bibinfo {year} {2022})}\BibitemShut {NoStop}%
\bibitem [{\citenamefont {Wouters}\ \emph {et~al.}(2008)\citenamefont {Wouters}, \citenamefont {Carusotto},\ and\ \citenamefont {Ciuti}}]{PhysRevB.77.115340}%
  \BibitemOpen
  \bibfield  {author} {\bibinfo {author} {\bibfnamefont {M.}~\bibnamefont {Wouters}}, \bibinfo {author} {\bibfnamefont {I.}~\bibnamefont {Carusotto}}, \ and\ \bibinfo {author} {\bibfnamefont {C.}~\bibnamefont {Ciuti}},\ }\href {\doibase 10.1103/PhysRevB.77.115340} {\bibfield  {journal} {\bibinfo  {journal} {Phys. Rev. B}\ }\textbf {\bibinfo {volume} {77}},\ \bibinfo {pages} {115340} (\bibinfo {year} {2008})}\BibitemShut {NoStop}%
\bibitem [{\citenamefont {Baas}\ \emph {et~al.}(2008)\citenamefont {Baas}, \citenamefont {Lagoudakis}, \citenamefont {Richard}, \citenamefont {Andr\'e}, \citenamefont {Dang},\ and\ \citenamefont {Deveaud-Pl\'edran}}]{PhysRevLett.100.170401}%
  \BibitemOpen
  \bibfield  {author} {\bibinfo {author} {\bibfnamefont {A.}~\bibnamefont {Baas}}, \bibinfo {author} {\bibfnamefont {K.~G.}\ \bibnamefont {Lagoudakis}}, \bibinfo {author} {\bibfnamefont {M.}~\bibnamefont {Richard}}, \bibinfo {author} {\bibfnamefont {R.}~\bibnamefont {Andr\'e}}, \bibinfo {author} {\bibfnamefont {L.~S.}\ \bibnamefont {Dang}}, \ and\ \bibinfo {author} {\bibfnamefont {B.}~\bibnamefont {Deveaud-Pl\'edran}},\ }\href {\doibase 10.1103/PhysRevLett.100.170401} {\bibfield  {journal} {\bibinfo  {journal} {Phys. Rev. Lett.}\ }\textbf {\bibinfo {volume} {100}},\ \bibinfo {pages} {170401} (\bibinfo {year} {2008})}\BibitemShut {NoStop}%
\bibitem [{\citenamefont {Ohadi}\ \emph {et~al.}(2016)\citenamefont {Ohadi}, \citenamefont {Gregory}, \citenamefont {Freegarde}, \citenamefont {Rubo}, \citenamefont {Kavokin}, \citenamefont {Berloff},\ and\ \citenamefont {Lagoudakis}}]{PhysRevX.6.031032}%
  \BibitemOpen
  \bibfield  {author} {\bibinfo {author} {\bibfnamefont {H.}~\bibnamefont {Ohadi}}, \bibinfo {author} {\bibfnamefont {R.~L.}\ \bibnamefont {Gregory}}, \bibinfo {author} {\bibfnamefont {T.}~\bibnamefont {Freegarde}}, \bibinfo {author} {\bibfnamefont {Y.~G.}\ \bibnamefont {Rubo}}, \bibinfo {author} {\bibfnamefont {A.~V.}\ \bibnamefont {Kavokin}}, \bibinfo {author} {\bibfnamefont {N.~G.}\ \bibnamefont {Berloff}}, \ and\ \bibinfo {author} {\bibfnamefont {P.~G.}\ \bibnamefont {Lagoudakis}},\ }\href {\doibase 10.1103/PhysRevX.6.031032} {\bibfield  {journal} {\bibinfo  {journal} {Phys. Rev. X}\ }\textbf {\bibinfo {volume} {6}},\ \bibinfo {pages} {031032} (\bibinfo {year} {2016})}\BibitemShut {NoStop}%
\bibitem [{\citenamefont {Rayanov}\ \emph {et~al.}(2015)\citenamefont {Rayanov}, \citenamefont {Altshuler}, \citenamefont {Rubo},\ and\ \citenamefont {Flach}}]{PhysRevLett.114.193901}%
  \BibitemOpen
  \bibfield  {author} {\bibinfo {author} {\bibfnamefont {K.}~\bibnamefont {Rayanov}}, \bibinfo {author} {\bibfnamefont {B.~L.}\ \bibnamefont {Altshuler}}, \bibinfo {author} {\bibfnamefont {Y.~G.}\ \bibnamefont {Rubo}}, \ and\ \bibinfo {author} {\bibfnamefont {S.}~\bibnamefont {Flach}},\ }\href {\doibase 10.1103/PhysRevLett.114.193901} {\bibfield  {journal} {\bibinfo  {journal} {Phys. Rev. Lett.}\ }\textbf {\bibinfo {volume} {114}},\ \bibinfo {pages} {193901} (\bibinfo {year} {2015})}\BibitemShut {NoStop}%
\bibitem [{\citenamefont {Amo}\ and\ \citenamefont {Bloch}(2016)}]{Amo_2016}%
  \BibitemOpen
  \bibfield  {author} {\bibinfo {author} {\bibfnamefont {A.}~\bibnamefont {Amo}}\ and\ \bibinfo {author} {\bibfnamefont {J.}~\bibnamefont {Bloch}},\ }\href {\doibase https://doi.org/10.1016/j.crhy.2016.08.007} {\bibfield  {journal} {\bibinfo  {journal} {Comptes Rendus Physique}\ }\textbf {\bibinfo {volume} {17}},\ \bibinfo {pages} {934} (\bibinfo {year} {2016})}\BibitemShut {NoStop}%
\bibitem [{\citenamefont {Alyatkin}\ \emph {et~al.}(2025)\citenamefont {Alyatkin}, \citenamefont {Sitnik}, \citenamefont {Daníelsson}, \citenamefont {Kartashov}, \citenamefont {Töpfer}, \citenamefont {Sigurðsson},\ and\ \citenamefont {Lagoudakis}}]{Alyatkin_Penrose}%
  \BibitemOpen
  \bibfield  {author} {\bibinfo {author} {\bibfnamefont {S.}~\bibnamefont {Alyatkin}}, \bibinfo {author} {\bibfnamefont {K.}~\bibnamefont {Sitnik}}, \bibinfo {author} {\bibfnamefont {V.~K.}\ \bibnamefont {Daníelsson}}, \bibinfo {author} {\bibfnamefont {Y.~V.}\ \bibnamefont {Kartashov}}, \bibinfo {author} {\bibfnamefont {J.~D.}\ \bibnamefont {Töpfer}}, \bibinfo {author} {\bibfnamefont {H.}~\bibnamefont {Sigurðsson}}, \ and\ \bibinfo {author} {\bibfnamefont {P.~G.}\ \bibnamefont {Lagoudakis}},\ }\href {\doibase 10.1126/sciadv.adz2484} {\bibfield  {journal} {\bibinfo  {journal} {Science Advances}\ }\textbf {\bibinfo {volume} {11}},\ \bibinfo {pages} {eadz2484} (\bibinfo {year} {2025})},\ \Eprint {http://arxiv.org/abs/https://www.science.org/doi/pdf/10.1126/sciadv.adz2484} {https://www.science.org/doi/pdf/10.1126/sciadv.adz2484} \BibitemShut {NoStop}%
\bibitem [{\citenamefont {T{\"o}pfer}\ \emph {et~al.}(2020)\citenamefont {T{\"o}pfer}, \citenamefont {Sigurdsson}, \citenamefont {Pickup},\ and\ \citenamefont {Lagoudakis}}]{Töpfer2020}%
  \BibitemOpen
  \bibfield  {author} {\bibinfo {author} {\bibfnamefont {J.~D.}\ \bibnamefont {T{\"o}pfer}}, \bibinfo {author} {\bibfnamefont {H.}~\bibnamefont {Sigurdsson}}, \bibinfo {author} {\bibfnamefont {L.}~\bibnamefont {Pickup}}, \ and\ \bibinfo {author} {\bibfnamefont {P.~G.}\ \bibnamefont {Lagoudakis}},\ }\href {\doibase 10.1038/s42005-019-0271-0} {\bibfield  {journal} {\bibinfo  {journal} {Communications Physics}\ }\textbf {\bibinfo {volume} {3}},\ \bibinfo {pages} {2} (\bibinfo {year} {2020})}\BibitemShut {NoStop}%
\bibitem [{\citenamefont {Harrison}\ \emph {et~al.}(2022)\citenamefont {Harrison}, \citenamefont {Sigurdsson}, \citenamefont {Alyatkin}, \citenamefont {T\"opfer},\ and\ \citenamefont {Lagoudakis}}]{PhysRevApplied.17.024063}%
  \BibitemOpen
  \bibfield  {author} {\bibinfo {author} {\bibfnamefont {S.}~\bibnamefont {Harrison}}, \bibinfo {author} {\bibfnamefont {H.}~\bibnamefont {Sigurdsson}}, \bibinfo {author} {\bibfnamefont {S.}~\bibnamefont {Alyatkin}}, \bibinfo {author} {\bibfnamefont {J.}~\bibnamefont {T\"opfer}}, \ and\ \bibinfo {author} {\bibfnamefont {P.}~\bibnamefont {Lagoudakis}},\ }\href {\doibase 10.1103/PhysRevApplied.17.024063} {\bibfield  {journal} {\bibinfo  {journal} {Phys. Rev. Appl.}\ }\textbf {\bibinfo {volume} {17}},\ \bibinfo {pages} {024063} (\bibinfo {year} {2022})}\BibitemShut {NoStop}%
\bibitem [{\citenamefont {Harrison}\ \emph {et~al.}(2020)\citenamefont {Harrison}, \citenamefont {Sigurdsson},\ and\ \citenamefont {Lagoudakis}}]{PhysRevB.101.155402}%
  \BibitemOpen
  \bibfield  {author} {\bibinfo {author} {\bibfnamefont {S.~L.}\ \bibnamefont {Harrison}}, \bibinfo {author} {\bibfnamefont {H.}~\bibnamefont {Sigurdsson}}, \ and\ \bibinfo {author} {\bibfnamefont {P.~G.}\ \bibnamefont {Lagoudakis}},\ }\href {\doibase 10.1103/PhysRevB.101.155402} {\bibfield  {journal} {\bibinfo  {journal} {Phys. Rev. B}\ }\textbf {\bibinfo {volume} {101}},\ \bibinfo {pages} {155402} (\bibinfo {year} {2020})}\BibitemShut {NoStop}%
\bibitem [{\citenamefont {Gnusov}\ \emph {et~al.}(2024)\citenamefont {Gnusov}, \citenamefont {Baryshev}, \citenamefont {Sigur{\dj}sson}, \citenamefont {Sitnik}, \citenamefont {T{\"o}pfer}, \citenamefont {Alyatkin},\ and\ \citenamefont {Lagoudakis}}]{gnusov2024observation}%
  \BibitemOpen
  \bibfield  {author} {\bibinfo {author} {\bibfnamefont {I.}~\bibnamefont {Gnusov}}, \bibinfo {author} {\bibfnamefont {S.}~\bibnamefont {Baryshev}}, \bibinfo {author} {\bibfnamefont {H.}~\bibnamefont {Sigur{\dj}sson}}, \bibinfo {author} {\bibfnamefont {K.}~\bibnamefont {Sitnik}}, \bibinfo {author} {\bibfnamefont {J.~D.}\ \bibnamefont {T{\"o}pfer}}, \bibinfo {author} {\bibfnamefont {S.}~\bibnamefont {Alyatkin}}, \ and\ \bibinfo {author} {\bibfnamefont {P.~G.}\ \bibnamefont {Lagoudakis}},\ }\href@noop {} {\bibfield  {journal} {\bibinfo  {journal} {Optica}\ }\textbf {\bibinfo {volume} {11}},\ \bibinfo {pages} {1156} (\bibinfo {year} {2024})}\BibitemShut {NoStop}%
\bibitem [{\citenamefont {Cilibrizzi}\ \emph {et~al.}(2014)\citenamefont {Cilibrizzi}, \citenamefont {Askitopoulos}, \citenamefont {Silva}, \citenamefont {Bastiman}, \citenamefont {Clarke}, \citenamefont {Zajac}, \citenamefont {Langbein},\ and\ \citenamefont {Lagoudakis}}]{doi:10.1063/1.4901814}%
  \BibitemOpen
  \bibfield  {author} {\bibinfo {author} {\bibfnamefont {P.}~\bibnamefont {Cilibrizzi}}, \bibinfo {author} {\bibfnamefont {A.}~\bibnamefont {Askitopoulos}}, \bibinfo {author} {\bibfnamefont {M.}~\bibnamefont {Silva}}, \bibinfo {author} {\bibfnamefont {F.}~\bibnamefont {Bastiman}}, \bibinfo {author} {\bibfnamefont {E.}~\bibnamefont {Clarke}}, \bibinfo {author} {\bibfnamefont {J.~M.}\ \bibnamefont {Zajac}}, \bibinfo {author} {\bibfnamefont {W.}~\bibnamefont {Langbein}}, \ and\ \bibinfo {author} {\bibfnamefont {P.~G.}\ \bibnamefont {Lagoudakis}},\ }\href {\doibase 10.1063/1.4901814} {\bibfield  {journal} {\bibinfo  {journal} {Applied Physics Letters}\ }\textbf {\bibinfo {volume} {105}},\ \bibinfo {pages} {191118} (\bibinfo {year} {2014})}\BibitemShut {NoStop}%
\bibitem [{\citenamefont {Chen}\ \emph {et~al.}(2013)\citenamefont {Chen}, \citenamefont {Mazilu}, \citenamefont {Arita}, \citenamefont {Wright},\ and\ \citenamefont {Dholakia}}]{chen2013dynamics}%
  \BibitemOpen
  \bibfield  {author} {\bibinfo {author} {\bibfnamefont {M.}~\bibnamefont {Chen}}, \bibinfo {author} {\bibfnamefont {M.}~\bibnamefont {Mazilu}}, \bibinfo {author} {\bibfnamefont {Y.}~\bibnamefont {Arita}}, \bibinfo {author} {\bibfnamefont {E.~M.}\ \bibnamefont {Wright}}, \ and\ \bibinfo {author} {\bibfnamefont {K.}~\bibnamefont {Dholakia}},\ }\href@noop {} {\bibfield  {journal} {\bibinfo  {journal} {Optics letters}\ }\textbf {\bibinfo {volume} {38}},\ \bibinfo {pages} {4919} (\bibinfo {year} {2013})}\BibitemShut {NoStop}%
\bibitem [{\citenamefont {Mart\'{\i}n}\ \emph {et~al.}(2002)\citenamefont {Mart\'{\i}n}, \citenamefont {Aichmayr}, \citenamefont {Vi\~na},\ and\ \citenamefont {Andr\'e}}]{PhysRevLett.89.077402}%
  \BibitemOpen
  \bibfield  {author} {\bibinfo {author} {\bibfnamefont {M.~D.}\ \bibnamefont {Mart\'{\i}n}}, \bibinfo {author} {\bibfnamefont {G.}~\bibnamefont {Aichmayr}}, \bibinfo {author} {\bibfnamefont {L.}~\bibnamefont {Vi\~na}}, \ and\ \bibinfo {author} {\bibfnamefont {R.}~\bibnamefont {Andr\'e}},\ }\href {\doibase 10.1103/PhysRevLett.89.077402} {\bibfield  {journal} {\bibinfo  {journal} {Phys. Rev. Lett.}\ }\textbf {\bibinfo {volume} {89}},\ \bibinfo {pages} {077402} (\bibinfo {year} {2002})}\BibitemShut {NoStop}%
\bibitem [{\citenamefont {Shelykh}\ \emph {et~al.}(2004)\citenamefont {Shelykh}, \citenamefont {Malpuech}, \citenamefont {Kavokin}, \citenamefont {Kavokin},\ and\ \citenamefont {Bigenwald}}]{Shelykh_PRB2004}%
  \BibitemOpen
  \bibfield  {author} {\bibinfo {author} {\bibfnamefont {I.}~\bibnamefont {Shelykh}}, \bibinfo {author} {\bibfnamefont {G.}~\bibnamefont {Malpuech}}, \bibinfo {author} {\bibfnamefont {K.~V.}\ \bibnamefont {Kavokin}}, \bibinfo {author} {\bibfnamefont {A.~V.}\ \bibnamefont {Kavokin}}, \ and\ \bibinfo {author} {\bibfnamefont {P.}~\bibnamefont {Bigenwald}},\ }\href {\doibase 10.1103/PhysRevB.70.115301} {\bibfield  {journal} {\bibinfo  {journal} {Phys. Rev. B}\ }\textbf {\bibinfo {volume} {70}},\ \bibinfo {pages} {115301} (\bibinfo {year} {2004})}\BibitemShut {NoStop}%
\bibitem [{\citenamefont {Krizhanovskii}\ \emph {et~al.}(2006)\citenamefont {Krizhanovskii}, \citenamefont {Sanvitto}, \citenamefont {Shelykh}, \citenamefont {Glazov}, \citenamefont {Malpuech}, \citenamefont {Solnyshkov}, \citenamefont {Kavokin}, \citenamefont {Ceccarelli}, \citenamefont {Skolnick},\ and\ \citenamefont {Roberts}}]{Krizhanovskii_PRB2006}%
  \BibitemOpen
  \bibfield  {author} {\bibinfo {author} {\bibfnamefont {D.~N.}\ \bibnamefont {Krizhanovskii}}, \bibinfo {author} {\bibfnamefont {D.}~\bibnamefont {Sanvitto}}, \bibinfo {author} {\bibfnamefont {I.~A.}\ \bibnamefont {Shelykh}}, \bibinfo {author} {\bibfnamefont {M.~M.}\ \bibnamefont {Glazov}}, \bibinfo {author} {\bibfnamefont {G.}~\bibnamefont {Malpuech}}, \bibinfo {author} {\bibfnamefont {D.~D.}\ \bibnamefont {Solnyshkov}}, \bibinfo {author} {\bibfnamefont {A.}~\bibnamefont {Kavokin}}, \bibinfo {author} {\bibfnamefont {S.}~\bibnamefont {Ceccarelli}}, \bibinfo {author} {\bibfnamefont {M.~S.}\ \bibnamefont {Skolnick}}, \ and\ \bibinfo {author} {\bibfnamefont {J.~S.}\ \bibnamefont {Roberts}},\ }\href {\doibase 10.1103/PhysRevB.73.073303} {\bibfield  {journal} {\bibinfo  {journal} {Phys. Rev. B}\ }\textbf {\bibinfo {volume} {73}},\ \bibinfo {pages} {073303} (\bibinfo {year} {2006})}\BibitemShut {NoStop}%
\bibitem [{\citenamefont {Solnyshkov}\ \emph {et~al.}(2007)\citenamefont {Solnyshkov}, \citenamefont {Shelykh}, \citenamefont {Glazov}, \citenamefont {Malpuech}, \citenamefont {Amand}, \citenamefont {Renucci}, \citenamefont {Marie},\ and\ \citenamefont {Kavokin}}]{Solnyshkov2007}%
  \BibitemOpen
  \bibfield  {author} {\bibinfo {author} {\bibfnamefont {D.~D.}\ \bibnamefont {Solnyshkov}}, \bibinfo {author} {\bibfnamefont {I.~A.}\ \bibnamefont {Shelykh}}, \bibinfo {author} {\bibfnamefont {M.~M.}\ \bibnamefont {Glazov}}, \bibinfo {author} {\bibfnamefont {G.}~\bibnamefont {Malpuech}}, \bibinfo {author} {\bibfnamefont {T.}~\bibnamefont {Amand}}, \bibinfo {author} {\bibfnamefont {P.}~\bibnamefont {Renucci}}, \bibinfo {author} {\bibfnamefont {X.}~\bibnamefont {Marie}}, \ and\ \bibinfo {author} {\bibfnamefont {A.~V.}\ \bibnamefont {Kavokin}},\ }\href {\doibase 10.1134/S1063782607090138} {\bibfield  {journal} {\bibinfo  {journal} {Semiconductors}\ }\textbf {\bibinfo {volume} {41}},\ \bibinfo {pages} {1080} (\bibinfo {year} {2007})}\BibitemShut {NoStop}%
\bibitem [{\citenamefont {Ryzhov}\ \emph {et~al.}(2020)\citenamefont {Ryzhov}, \citenamefont {Kozlov}, \citenamefont {Kuznetsov}, \citenamefont {Chestnov}, \citenamefont {Kavokin}, \citenamefont {Tzimis}, \citenamefont {Hatzopoulos}, \citenamefont {Savvidis}, \citenamefont {Kozlov},\ and\ \citenamefont {Zapasskii}}]{Ryzhov_PhysRevRes2020}%
  \BibitemOpen
  \bibfield  {author} {\bibinfo {author} {\bibfnamefont {I.~I.}\ \bibnamefont {Ryzhov}}, \bibinfo {author} {\bibfnamefont {V.~O.}\ \bibnamefont {Kozlov}}, \bibinfo {author} {\bibfnamefont {N.~S.}\ \bibnamefont {Kuznetsov}}, \bibinfo {author} {\bibfnamefont {I.~Y.}\ \bibnamefont {Chestnov}}, \bibinfo {author} {\bibfnamefont {A.~V.}\ \bibnamefont {Kavokin}}, \bibinfo {author} {\bibfnamefont {A.}~\bibnamefont {Tzimis}}, \bibinfo {author} {\bibfnamefont {Z.}~\bibnamefont {Hatzopoulos}}, \bibinfo {author} {\bibfnamefont {P.~G.}\ \bibnamefont {Savvidis}}, \bibinfo {author} {\bibfnamefont {G.~G.}\ \bibnamefont {Kozlov}}, \ and\ \bibinfo {author} {\bibfnamefont {V.~S.}\ \bibnamefont {Zapasskii}},\ }\href {\doibase 10.1103/PhysRevResearch.2.022064} {\bibfield  {journal} {\bibinfo  {journal} {Phys. Rev. Research}\ }\textbf {\bibinfo {volume} {2}},\ \bibinfo {pages} {022064} (\bibinfo {year} {2020})}\BibitemShut {NoStop}%
\bibitem [{\citenamefont {del Valle-Inclan~Redondo}\ \emph {et~al.}(2019{\natexlab{b}})\citenamefont {del Valle-Inclan~Redondo}, \citenamefont {Sigurdsson}, \citenamefont {Ohadi}, \citenamefont {Shelykh}, \citenamefont {Rubo}, \citenamefont {Hatzopoulos}, \citenamefont {Savvidis},\ and\ \citenamefont {Baumberg}}]{PhysRevB.99.165311}%
  \BibitemOpen
  \bibfield  {author} {\bibinfo {author} {\bibfnamefont {Y.}~\bibnamefont {del Valle-Inclan~Redondo}}, \bibinfo {author} {\bibfnamefont {H.}~\bibnamefont {Sigurdsson}}, \bibinfo {author} {\bibfnamefont {H.}~\bibnamefont {Ohadi}}, \bibinfo {author} {\bibfnamefont {I.~A.}\ \bibnamefont {Shelykh}}, \bibinfo {author} {\bibfnamefont {Y.~G.}\ \bibnamefont {Rubo}}, \bibinfo {author} {\bibfnamefont {Z.}~\bibnamefont {Hatzopoulos}}, \bibinfo {author} {\bibfnamefont {P.~G.}\ \bibnamefont {Savvidis}}, \ and\ \bibinfo {author} {\bibfnamefont {J.~J.}\ \bibnamefont {Baumberg}},\ }\href {\doibase 10.1103/PhysRevB.99.165311} {\bibfield  {journal} {\bibinfo  {journal} {Phys. Rev. B}\ }\textbf {\bibinfo {volume} {99}},\ \bibinfo {pages} {165311} (\bibinfo {year} {2019}{\natexlab{b}})}\BibitemShut {NoStop}%
\bibitem [{\citenamefont {Sigurdsson}\ \emph {et~al.}(2022)\citenamefont {Sigurdsson}, \citenamefont {Gnusov}, \citenamefont {Alyatkin}, \citenamefont {Pickup}, \citenamefont {Gippius}, \citenamefont {Lagoudakis},\ and\ \citenamefont {Askitopoulos}}]{PhysRevLett.129.155301}%
  \BibitemOpen
  \bibfield  {author} {\bibinfo {author} {\bibfnamefont {H.}~\bibnamefont {Sigurdsson}}, \bibinfo {author} {\bibfnamefont {I.}~\bibnamefont {Gnusov}}, \bibinfo {author} {\bibfnamefont {S.}~\bibnamefont {Alyatkin}}, \bibinfo {author} {\bibfnamefont {L.}~\bibnamefont {Pickup}}, \bibinfo {author} {\bibfnamefont {N.~A.}\ \bibnamefont {Gippius}}, \bibinfo {author} {\bibfnamefont {P.~G.}\ \bibnamefont {Lagoudakis}}, \ and\ \bibinfo {author} {\bibfnamefont {A.}~\bibnamefont {Askitopoulos}},\ }\href {\doibase 10.1103/PhysRevLett.129.155301} {\bibfield  {journal} {\bibinfo  {journal} {Phys. Rev. Lett.}\ }\textbf {\bibinfo {volume} {129}},\ \bibinfo {pages} {155301} (\bibinfo {year} {2022})}\BibitemShut {NoStop}%
\bibitem [{\citenamefont {Gnusov}\ \emph {et~al.}(2025)\citenamefont {Gnusov}, \citenamefont {Yulin}, \citenamefont {Baryshev}, \citenamefont {Alyatkin},\ and\ \citenamefont {Lagoudakis}}]{lxv5-qwxw}%
  \BibitemOpen
  \bibfield  {author} {\bibinfo {author} {\bibfnamefont {I.}~\bibnamefont {Gnusov}}, \bibinfo {author} {\bibfnamefont {A.}~\bibnamefont {Yulin}}, \bibinfo {author} {\bibfnamefont {S.}~\bibnamefont {Baryshev}}, \bibinfo {author} {\bibfnamefont {S.}~\bibnamefont {Alyatkin}}, \ and\ \bibinfo {author} {\bibfnamefont {P.~G.}\ \bibnamefont {Lagoudakis}},\ }\href {\doibase 10.1103/lxv5-qwxw} {\bibfield  {journal} {\bibinfo  {journal} {Phys. Rev. B}\ }\textbf {\bibinfo {volume} {112}},\ \bibinfo {pages} {L201302} (\bibinfo {year} {2025})}\BibitemShut {NoStop}%
\bibitem [{\citenamefont {Dovzhenko}\ \emph {et~al.}(2023)\citenamefont {Dovzhenko}, \citenamefont {Aristov}, \citenamefont {Pickup}, \citenamefont {Sigur{\dh}sson},\ and\ \citenamefont {Lagoudakis}}]{dovzhenko2023next}%
  \BibitemOpen
  \bibfield  {author} {\bibinfo {author} {\bibfnamefont {D.}~\bibnamefont {Dovzhenko}}, \bibinfo {author} {\bibfnamefont {D.}~\bibnamefont {Aristov}}, \bibinfo {author} {\bibfnamefont {L.}~\bibnamefont {Pickup}}, \bibinfo {author} {\bibfnamefont {H.}~\bibnamefont {Sigur{\dh}sson}}, \ and\ \bibinfo {author} {\bibfnamefont {P.}~\bibnamefont {Lagoudakis}},\ }\href@noop {} {\bibfield  {journal} {\bibinfo  {journal} {Physical Review B}\ }\textbf {\bibinfo {volume} {108}},\ \bibinfo {pages} {L161301} (\bibinfo {year} {2023})}\BibitemShut {NoStop}%
\bibitem [{\citenamefont {Gnusov}\ \emph {et~al.}(2021)\citenamefont {Gnusov}, \citenamefont {Sigurdsson}, \citenamefont {T\"opfer}, \citenamefont {Baryshev}, \citenamefont {Alyatkin},\ and\ \citenamefont {Lagoudakis}}]{gnusov_prapl}%
  \BibitemOpen
  \bibfield  {author} {\bibinfo {author} {\bibfnamefont {I.}~\bibnamefont {Gnusov}}, \bibinfo {author} {\bibfnamefont {H.}~\bibnamefont {Sigurdsson}}, \bibinfo {author} {\bibfnamefont {J.}~\bibnamefont {T\"opfer}}, \bibinfo {author} {\bibfnamefont {S.}~\bibnamefont {Baryshev}}, \bibinfo {author} {\bibfnamefont {S.}~\bibnamefont {Alyatkin}}, \ and\ \bibinfo {author} {\bibfnamefont {P.}~\bibnamefont {Lagoudakis}},\ }\href {\doibase 10.1103/PhysRevApplied.16.034014} {\bibfield  {journal} {\bibinfo  {journal} {Phys. Rev. Applied}\ }\textbf {\bibinfo {volume} {16}},\ \bibinfo {pages} {034014} (\bibinfo {year} {2021})}\BibitemShut {NoStop}%
\end{thebibliography}
\end{document}

\title{Supplementary Materials: Non-local Stabilization and Synchronization of Larmor Precession in Trapped Polariton Condensate Dimer}

\maketitle

\section{theory}

\subsection{Coupled mode approach and Kuramoto model}

To describe the dynamics of the condensates we adopt a simple and effective model based on coupled modes theory. We assume that the condensate can be represented as a set of interacting modes of different angular momenta and polarizations. This method is proven to be efficient tool of description for the dynamics of the condensates provided that the inter-mode interactions are sufficiently weak, see \cite{lagoudakis2017polariton,Berloff2017,chestnov2019optical}. The derivation of the equation for the case of two interacting condensate is given below. 

We assume that the parameters of the system is such that the incoherent pump excites only the fundamental mode with zero angular momentum. It is convenient to describe the dynamics the condensate in the basis of circular polarizations. The existence of the axially non-uniform pump couples the modes of different polarizations via non-resonant excitation of the angular-dependent field. We assume that the interaction between the modes is so weak that it does not affect much the spatial distribution of the fields and can be accounted by the complex amplitudes of the modes. As it is discussed in detail in \cite{lxv5-qwxw} the equations for the modes amplitudes read 
\begin{eqnarray}
    \dot A_{1 \uparrow}=(-\gamma +\frac{(1+i\alpha) P_{1 \uparrow}}{1+|A_{1 \uparrow}|^2} +i\mu |A_{1 \uparrow}|^2)A_{1 \uparrow} + (1+i\alpha)\sigma e^{2i\Omega t}A_{1\downarrow}   \label{eq_A_first_c_1}  \\
    \dot A_{1 \downarrow}=(-\gamma +\frac{(1+i\alpha) P_{1 \downarrow}}{1+|A_{1 \downarrow}|^2} +i\mu |A_{1 \downarrow}|^2)A_{1 \downarrow} + (1+i\alpha)\sigma e^{-2i\Omega t} A_{1\uparrow}  \label{eq_A_first_c_2}  
\end{eqnarray}
where subscripts $\uparrow$ and $\downarrow$ denote clock- and counter-clockwise polarizations, subindex $1$ shows that the quantity relates to the condensate $C_1$ driven by a rotating potential. The amplitudes of the polarizations are $A_{1 \uparrow}$ and $A_{1 \downarrow}$, $\gamma$ is the decay rate of the polaritons, $\alpha$ is the ratio of the blueshift to the gain caused by the reservoir of incoherent excitons, $\mu$ accounts for the blue shift produced by polariton-polariton interaction. 

The axially non-symmetric pump results in the coupling between the polarizations, the coupling strength is $\sigma$. The pump rotates with the angular velocity $\Omega$ and therefore the effective coupling between the modes depends on time. Let us remark that in the reference frame rotating at the velocity $\Omega$ the pump becomes time-independent and this is why the explicit dependency of the coupling on time can be removed by the variables change $A_{1 \uparrow} = B_{1 \uparrow} e^{i\Omega t}$, $A_{1 \downarrow} = B_{1 \downarrow} e^{-i\Omega t}$. The parameters $\gamma$, $\alpha$ and $\mu$ can be found by fitting the experimental dependencies of the condensate frequency on the pump intensity \cite{lxv5-qwxw}. 

The equations governing the dynamics of the $C_2$ condensate are the same but the absence of the axially non-symmetric potential removes the coupling between the polarizations
\begin{eqnarray}
    \dot A_{2 \uparrow}=(-\gamma +\frac{(1+i\alpha) P_{2 \uparrow}}{1+|A_{2 \uparrow}|^2} +i\mu |A_{2 \uparrow}|^2)A_{2 \uparrow}    \label{eq_A_second_c_1}  \\
    \dot A_{2 \downarrow}=(-\gamma +\frac{(1+i\alpha) P_{2 \downarrow}}{1+|A_{2 \downarrow}|^2} +i\mu |A_{2 \downarrow}|^2)A_{2 \downarrow},  \label{eq_A_second_c_2}  
\end{eqnarray}
subindex $2$ shows that the quantities relate to the condensate $C_2$.

Now we make the next step and introduce the coupling between $C_2$ and $C_1$ condensates. The coupling appears because the polariton field of $C_1$ and $C_2$ condensates overlap. Let us remark that this can be the overlap of the evanescent fields or the field of ballistic polaritons propagating from one condensate to another. Regardless of the exact mechanism of the interaction between the condensates, in our model the interaction between the condensates can be accounted by some constants $\epsilon$ and $\tilde \epsilon$. This approximation works provided that the distance between the condensates is so small that the interaction between them can be considered as instantaneous (the characteristic time of the condensates dynamics is much longer than the time needed for polaritons to travel between the condensates). 

Two cross-condensates interaction constants are needed because the neighboring condensate affect both the eigenfrequency (this is accounted by $\tilde \epsilon =\tilde \epsilon_0\left(1+q\cos (2\Omega t)\right)$) and the effective losses (accounted by $\epsilon=\epsilon_0\left(1+q\cos (2\Omega t)\right)$) experienced by the condensate. Let us remark that the angular-dependent pump make the transparency of the effective potential trapping the $C_1$ condensate angular dependent. Therefore the effective coupling between the condensates depend on the orientation on the pump in respect to the axis passing through the centres of the condensates. The rotation of the pump makes the coupling strength time dependent, the modulation of the coupling strength is defined by $q<1$ and the frequency of the coupling strength modulation is exactly equal to the doubled  angular velocity of the potential. In the experiments the pump sinusoidally modulated along the angular coordinate $\sim \sin(2\theta)$ is used. For such shape of the pump the effective coupling changes sinusoidally provided that the pump modulation is shallow.

Finally, we can write the equations for the amplitudes of the condensates in the form
\begin{eqnarray}
    \dot A_{1 \uparrow}=(-\gamma +\frac{(1+i\alpha) P_{1 \uparrow}}{1+|A_{1 \uparrow}|^2} +i\mu |A_{1 \uparrow}|^2)A_{1 \uparrow} + (1+i\alpha)\sigma e^{2i\Omega t}A_{1\downarrow} +(\epsilon+i\tilde \epsilon)  A_{2 \uparrow} \label{eq_A_master_cL_p_up} \\
    \dot A_{1 \downarrow}=(-\gamma +\frac{(1+i\alpha) P_{1 \downarrow}}{1+|A_{1 \downarrow}|^2} +i\mu |A_{1 \downarrow}|^2)A_{1 \downarrow} + (1+i\alpha)\sigma e^{-2i\Omega t} A_{1\uparrow} +(\epsilon+i\tilde \epsilon)   A_{2 \downarrow} \label{eq_A_master_cL_p_down} \\
\dot A_{2 \uparrow}=(-\gamma +\frac{(1+i\alpha) P_{2 \uparrow}}{1+|A_{2 \uparrow}|^2} +i\mu |A_{2 \uparrow}|^2)A_{2 \uparrow}  +(\epsilon+i\tilde \epsilon)  A_{1 \uparrow} \label{eq_A_master_cR_p_up} \\
\dot A_{2 \downarrow}=(-\gamma +\frac{(1+i\alpha) P_{2 \downarrow}}{1+|A_{2 \downarrow}|^2} +i\mu |A_{2 \downarrow}|^2)A_{2 \downarrow}  +(\epsilon+i\tilde \epsilon)   A_{1 \downarrow} \label{eq_A_master_cR_p_down}. 
\end{eqnarray}

To proceed we look for the solution in the form $A_{1,2 \, \uparrow,\downarrow}=\rho_{1,2 \,  \uparrow,\downarrow} e^{i\omega_{1,2 \, \uparrow,\downarrow} t +i \phi_{1,2 \, \uparrow,\downarrow}}$ where $\rho$ are the real amplitudes and $\phi$ are the phases of the modes, $\omega$ - their free-running frequencies (the frequency of the modes not perturbed by their mutual interaction or by the rotating pump ). 

Substituting this into (\ref{eq_A_master_cL_p_up})-(\ref{eq_A_master_cR_p_down}) and separating the real and imaginary parts we obtain the equations for $\rho$ and $\phi$
\begin{eqnarray}
    \dot \rho_{1 \uparrow} =\left( -\gamma +\frac{P_{1 \uparrow}}{1+\rho_{1 \uparrow}^2} \right)\rho_{1 \uparrow}+\sigma \rho_{1 \downarrow} \left( \cos\left( \delta t - \phi_{1 \downarrow}+\phi_{1 \uparrow} \right)+\alpha   \sin\left( \delta t - \phi_{1 \downarrow}+\phi_{1 \uparrow} \right) \right) \nonumber \\
    +\rho_{2 \uparrow} \left( \epsilon \cos\left( \delta_{\uparrow}t - \phi_{2 \uparrow}+\phi_{1 \uparrow} \right) + \tilde \epsilon \sin\left( \delta_{\uparrow}t - \phi_{2 \uparrow}+\phi_{1 \uparrow} \right)  \right),  \label{eq_rho_cL_pUp} \\
    \dot \rho_{1 \downarrow} =\left( -\gamma +\frac{P_{1 \downarrow}}{1+\rho_{1 \downarrow}^2} \right)\rho_{1 \downarrow}+\sigma \rho_{1 \uparrow} \left( \cos\left( \delta t + \phi_{1 \uparrow}-\phi_{1 \downarrow} \right)-\alpha   \sin\left( \delta t + \phi_{1 \uparrow}-\phi_{1 \downarrow} \right) \right) \nonumber \\
    +\rho_{2 \downarrow}  \left( \epsilon \cos\left( \delta_{\downarrow}t - \phi_{2 \downarrow}+\phi_{1 \downarrow} \right) + \tilde \epsilon \sin\left( \delta_{\downarrow}t - \phi_{2 \downarrow}+\phi_{1 \downarrow} \right)  \right), \label{eq_rho_cL_pDown} \\
    \dot \rho_{2 \uparrow} =\left( -\gamma +\frac{P_{2 \uparrow}}{1+\rho_{2 \uparrow}^2} \right)\rho_{2 \uparrow} 
    +\rho_{1 \uparrow} \left( \epsilon \cos\left( \delta_{\uparrow} t + \phi_{1 \uparrow}-\phi_{2 \uparrow} \right) - \tilde \epsilon \sin\left( \delta_{\uparrow} t + \phi_{1 \uparrow}-\phi_{2 \uparrow} \right)  \right), \label{eq_rho_cR_pUp} \\
    \dot \rho_{2 \downarrow} =\left( -\gamma +\frac{P_{2 \downarrow}}{1+\rho_{2 \downarrow}^2} \right)\rho_{2 \downarrow} 
    +\rho_{1 \downarrow}  \left( \epsilon \cos\left( \delta_{\downarrow}t + \phi_{1 \downarrow}-\phi_{2 \downarrow} \right) - \tilde \epsilon \sin\left( \delta_{\downarrow}t + \phi_{1 \downarrow}-\phi_{2 \downarrow} \right)  \right), \label{eq_rho_cR_pDown} \\
    \dot \phi_{1 \uparrow}  =\left(\frac{\alpha P_{1 \uparrow}}{1+\rho_{1 \uparrow}^2} +\mu \rho_{1 \uparrow}^2 -\omega_{1 \uparrow} \right)-\sigma \frac{\rho_{1 \downarrow}}{ \rho_{1 \uparrow}} \left( \sin\left( \delta t - \phi_{1 \downarrow}+\phi_{1 \uparrow} \right)-\alpha   \cos\left( \delta t - \phi_{1 \downarrow}+\phi_{1 \uparrow} \right) \right) \nonumber \\
    -\frac{\rho_{2 \uparrow} } { \rho_{1 \uparrow}}\left(  \epsilon \sin \left( \delta_{\uparrow}t - \phi_{2 \uparrow}+\phi_{1 \uparrow} \right) - \tilde \epsilon \cos\left( \delta_{\uparrow}t - \phi_{2 \uparrow}+\phi_{1 \uparrow} \right)  \right), \label{eq_phi_cL_pUp} \\
    \dot \phi_{1 \downarrow}  =\left( \frac{\alpha P_{1 \downarrow}}{1+\rho_{1 \downarrow}^2} +\mu \rho_{1 \downarrow}^2 -\omega_{1 \downarrow} \right)+\sigma \frac{\rho_{1 \uparrow}}{\rho_{1 \downarrow}} \left( \sin \left( \delta t + \phi_{1 \uparrow}-\phi_{1 \downarrow} \right)+\alpha   \cos \left( \delta t + \phi_{1 \uparrow}-\phi_{1 \downarrow} \right) \right) \nonumber \\
    -\frac{\rho_{2 \downarrow} }{\rho_{1 \downarrow}}\left( \epsilon \sin \left( \delta_{\downarrow}t - \phi_{2 \downarrow}+\phi_{1 \downarrow} \right) - \tilde \epsilon \cos \left( \delta_{\downarrow}t - \phi_{2 \downarrow}+\phi_{1 \downarrow} \right)  \right), \label{eq_phi_cL_pDown} \\
    \dot \phi_{2 \uparrow}  =\left( \frac{\alpha P_{2 \uparrow}}{1+\rho_{2 \uparrow}^2} + \mu \rho_{2 \uparrow}^2 - \omega_{2 \uparrow} \right)
    +\frac{\rho_{1 \uparrow} }{\rho_{2 \uparrow}}\left( \epsilon \sin\left( \delta_{\uparrow} t + \phi_{1 \uparrow}-\phi_{2 \uparrow} \right) + \tilde \epsilon \cos\left( \delta_{\uparrow} t + \phi_{1 \uparrow}-\phi_{2 \uparrow} \right)  \right), \label{eq_phi_cR_pUp} \\
    \dot \phi_{2 \downarrow}  =\left( \frac{\alpha P_{2 \downarrow}}{1+\rho_{2 \downarrow}^2}  +\mu \rho_{2 \downarrow}^2 - \omega_{2 \downarrow} \right) 
    +\frac{\rho_{1 \downarrow}}{\rho_{2 \downarrow}} \left( \epsilon \sin\left( \delta_{\downarrow}t + \phi_{1 \downarrow}-\phi_{2 \downarrow} \right) + \tilde \epsilon \cos\left( \delta_{\downarrow}t + \phi_{1 \downarrow}-\phi_{2 \downarrow} \right)  \right) \label{eq_phi_cR_pDown}.    
\end{eqnarray}
The amplitudes  of the non-interacting modes are $\bar \rho_{1 \uparrow}=\frac{P_{1 \uparrow}}{\gamma}-1$, $\bar \rho_{2 \uparrow}=\frac{P_{2 \uparrow}}{\gamma}-1$, $\bar \rho_{1 \downarrow}=\frac{P_{1 \downarrow}}{\gamma}-1$, $\bar \rho_{2 \downarrow}=\frac{P_{2 \downarrow}}{\gamma}-1$ and, then the free-running frequencies are $\omega_{1 \uparrow}=\alpha \gamma  +\mu \rho_{1 \uparrow}^2$, $\omega_{1 \downarrow}=\alpha \gamma  +\mu \rho_{1 \downarrow}^2$, $\omega_{2 \uparrow}=\alpha \gamma  +\mu \rho_{2 \uparrow}^2$, $\omega_{2 \downarrow}=\alpha \gamma  +\mu \rho_{2 \downarrow}^2$. 

Further we will be interested in the dynamics of nearly equally pumped condensates  $\frac{P_{{1 \uparrow}}}{\gamma}-1 \approx \frac{P_{{1 \downarrow}}}{\gamma}-1 \approx \frac{P_{{2 \uparrow}}}{\gamma}-1 \approx \frac{P_{{2 \downarrow}}}{\gamma}-1 = \frac{P}{\gamma}-1$ so that the amplitudes of the modes are similar $\rho_{1 \uparrow} \approx \rho_{1 \downarrow}\approx \rho_{2 \uparrow} \approx \rho_{2 \downarrow} = \rho=\frac{P}{\gamma}-1$.  Then the conditions that the coupling between the modes does not affect much their amplitudes  is $\epsilon_0, \tilde \epsilon_0, \sigma, \alpha \sigma \ll \Gamma$, where 
$\Gamma=2\gamma \left( 1-\frac{\gamma}{P} \right) $ is the amplitudes relaxation rate. For pump powers well above threshold, the relaxation rate $\Gamma$ can be estimated as twice the linear polariton relaxation rate, yielding an approximate value of $100\text{ ps}^{-1}$.

The condition of  weak coupling allows us to look for the solution for the real amplitudes $\rho$ as $\rho=\bar \rho + r$ where $r \ll \rho$ accounts for the effect of the coupling. To proceed we assume that the phases $\phi$ vary slowly in the sense that $| \dot \phi| \ll \Gamma $. A posteriori we define the condition that the assumption is met. The discussed assumption make it possible to neglect the derivatives in (\ref{eq_rho_cL_pUp})-(\ref{eq_rho_cR_pDown}) and reduce them to the algebraic equation linear in respect to $r$. Then the solution for the perturbation of the mode amplitudes are
\begin{eqnarray}
    r_{1 \uparrow} = \frac{\sigma \rho}{\Gamma} \left(  \cos\left( \delta t - \phi_{1 \downarrow}+\phi_{1 \uparrow} \right)+\alpha  \sin\left( \delta t - \phi_{1 \downarrow} + \phi_{1 \uparrow} \right)\right) +\nonumber \\ 
     +\frac{\rho}{\Gamma}  \left(\epsilon \cos\left( \delta_{\uparrow}t - \phi_{2 \uparrow}+\phi_{1 \uparrow} \right) + \tilde \epsilon \sin\left( \delta_{\uparrow}t - \phi_{2 \uparrow}+\phi_{1 \uparrow} \right) \right) , \label{eq_r_cL_pUp} \\
    r_{1 \downarrow} =\frac{\sigma \rho}{\Gamma} \left(  \sigma \cos\left( \delta t + \phi_{1 \uparrow}-\phi_{1 \downarrow} \right)-\alpha \sigma  \sin\left( \delta t + \phi_{1 \uparrow}-\phi_{1 \downarrow} \right) \right)+  \nonumber \\
    + \frac{\rho}{\Gamma}\left( \epsilon \cos\left( \delta_{\downarrow}t - \phi_{2 \downarrow}+\phi_{1 \downarrow} \right) + \tilde \epsilon \sin\left( \delta_{\downarrow}t - \phi_{2 \downarrow}+\phi_{1 \downarrow} \right) \right) , \label{eq_r_cL_pDown}\\
   r_{2 \uparrow} =\frac{\rho}{\Gamma}  \left( \epsilon \cos\left( \delta_{\uparrow} t + \phi_{1 \uparrow}-\phi_{2 \uparrow} \right) - \tilde \epsilon \sin\left( \delta_{\uparrow} t + \phi_{1 \uparrow}-\phi_{2 \uparrow} \right)  \right), \label{eq_r_cR_pUp} \\
   r_{2 \downarrow} =\frac{\rho}{\Gamma}  \left( \epsilon \cos\left( \delta_{\downarrow}t + \phi_{1 \downarrow}-\phi_{2 \downarrow} \right) - \tilde \epsilon \sin\left( \delta_{\downarrow}t + \phi_{1 \downarrow}-\phi_{2 \downarrow} \right)  \right) \label{eq_r_cR_pDown}.  
\end{eqnarray}

Using the condition that the variations of the modes amplitudes are small we expand nonlinear terms in Taylor series and rewrite the equations (\ref{eq_phi_cL_pUp})-(\ref{eq_phi_cR_pDown}) for the phases keeping only terms linear in respect to $r$
\begin{eqnarray}
    \dot \phi_{1 \uparrow}  = 2\rho \left( \mu-\frac{\alpha \gamma^2}{P}  \right) r_{1 \uparrow}-\sigma  \left( \sin\left( \delta t - \phi_{1 \downarrow}+\phi_{1 \uparrow} \right)-\alpha   \cos\left( \delta t - \phi_{1 \downarrow}+\phi_{1 \uparrow} \right) \right) \nonumber \\
    -\left(  \epsilon \sin \left( \delta_{\uparrow}t - \phi_{2 \uparrow}+\phi_{1 \uparrow} \right) - \tilde \epsilon \cos\left( \delta_{\uparrow}t - \phi_{2 \uparrow}+\phi_{1 \uparrow} \right)  \right),  \label{eq_phi_cL_pUp_} \\
    \dot \phi_{1 \downarrow}  =2\rho \left( \mu-\frac{\alpha \gamma^2}{P}  \right) r_{1 \downarrow}+\sigma  \left( \sin \left( \delta t + \phi_{1 \uparrow}-\phi_{1 \downarrow} \right)+\alpha   \cos \left( \delta t + \phi_{1 \uparrow}-\phi_{1 \downarrow} \right) \right) \nonumber \\
    -\left( \epsilon \sin \left( \delta_{\downarrow}t - \phi_{2 \downarrow}+\phi_{1 \downarrow} \right) - \tilde \epsilon \cos \left( \delta_{\downarrow}t - \phi_{2 \downarrow}+\phi_{1 \downarrow} \right)  \right), \label{eq_phi_cL_pDown_} \\
    \dot \phi_{2 \uparrow}  =2\rho \left( \mu-\frac{\alpha \gamma^2}{P}  \right) r_{2 \uparrow}
    +\left( \epsilon \sin\left( \delta_{\uparrow} t + \phi_{1 \uparrow}-\phi_{2 \uparrow} \right) + \tilde \epsilon \cos\left( \delta_{\uparrow} t + \phi_{1 \uparrow}-\phi_{2 \uparrow} \right)  \right), \label{eq_phi_cR_pUp_} \\
    \dot \phi_{2 \downarrow}  =2\rho \left( \mu-\frac{\alpha \gamma^2}{P}  \right) r_{2 \downarrow} 
    +  \left( \epsilon \sin\left( \delta_{\downarrow}t + \phi_{1 \downarrow}-\phi_{2 \downarrow} \right) + \tilde \epsilon \cos\left( \delta_{\downarrow}t + \phi_{1 \downarrow}-\phi_{2 \downarrow} \right)  \right) \label{eq_phi_cR_pDown_}.    
\end{eqnarray}

Finally we substitute (\ref{eq_r_cL_pUp})-(\ref{eq_r_cR_pDown} ) into (\ref{eq_phi_cR_pDown_})-(\ref{eq_phi_cR_pDown_}) and obtain the equations for the phases
\begin{eqnarray}
    \dot \phi_{1 \uparrow}  = \sigma  \left(  \frac{\mu P}{\gamma^2} 
        \cos\left( \delta t - \phi_{1 \downarrow}+\phi_{1 \uparrow} \right)+\left(  \frac{\alpha \mu P}{\gamma^2}-\alpha^2  - 1 \right)
        \sin\left( \delta t - \phi_{1 \downarrow}+\phi_{1 \uparrow} \right) \right)+ \nonumber \\
        + \left( \left( \frac{\mu P}{\gamma^2}-\alpha \right)\epsilon  +\tilde \epsilon \right)
        \cos\left( \delta_{\uparrow} t - \phi_{2 \uparrow}+\phi_{1 \uparrow} \right)+\left( \left(  \frac{\mu P}{\gamma^2}-\alpha \right) \tilde \epsilon - \epsilon \right)
        \sin\left( \delta_{\uparrow} t - \phi_{2 \uparrow}+\phi_{1 \uparrow} \right)  \label{eq_phase_cL_pUp} \\
    \dot \phi_{1 \downarrow}  = \sigma  \left(  \frac{\mu P}{\gamma^2} 
        \cos\left( \delta t - \phi_{1 \downarrow}+\phi_{1 \uparrow} \right)+\left(  \frac{\alpha \mu P}{\gamma^2}-\alpha^2  - 1 \right)
        \sin\left( \delta t - \phi_{1 \uparrow}+\phi_{1 \downarrow} \right) \right)+ \nonumber \\
        + \left( \left( \frac{\mu P}{\gamma^2}-\alpha \right)\epsilon  +\tilde \epsilon \right)
        \cos\left( \delta_{\downarrow} t - \phi_{2 \downarrow}+\phi_{1 \downarrow} \right)+\left( \left(  \frac{\mu P}{\gamma^2}-\alpha \right) \tilde \epsilon - \epsilon \right)
        \sin\left( \delta_{\downarrow} t - \phi_{2 \downarrow}+\phi_{1 \downarrow} \right)      \label{eq_phase_cL_pDown}\\
    \dot \phi_{2 \uparrow}  = 
        \left( \left( \frac{\mu P}{\gamma^2}-\alpha \right)\epsilon  +\tilde \epsilon \right)
        \cos\left( \delta_{\uparrow} t - \phi_{2 \uparrow}+\phi_{1 \uparrow} \right)-\left( \left(  \frac{\mu P}{\gamma^2}-\alpha \right) \tilde \epsilon - \epsilon \right)
        \sin\left( \delta_{\uparrow} t - \phi_{2 \uparrow}+\phi_{1 \uparrow} \right)   \label{eq_phase_cR_pUp} \\
    \dot \phi_{2 \downarrow}  = 
         \left( \left( \frac{\mu P}{\gamma^2}-\alpha \right)\epsilon  +\tilde \epsilon \right)
        \cos\left( \delta_{\downarrow} t - \phi_{2 \downarrow}+\phi_{1 \downarrow} \right)-\left( \left(  \frac{\mu P}{\gamma^2}-\alpha \right) \tilde \epsilon - \epsilon \right)
        \sin\left( \delta_{\downarrow} t - \phi_{2 \downarrow}+\phi_{1 \downarrow} \right)   \label{eq_phase_cR_pDown}.              
\end{eqnarray}

It is convenient to introduce the relative phases defined as $\phi=\phi_{1 \, \uparrow} - \phi_{1 \, \downarrow}  + \delta t $, $\phi_{\uparrow}=\phi_{1 \, \uparrow} - \phi_{2 \, \uparrow}  + \delta_{\uparrow}  t$ and $\phi_{\downarrow}=\phi_{1 \, \downarrow} - \phi_{2 \, \downarrow}  + \delta_{\downarrow}  t$ where 
$\delta=\omega_{1 \, \uparrow}- \omega_{1 \, \downarrow}-2\Omega $
$\delta_{\uparrow}=\omega_{1 \, \uparrow}- \omega_{2 \, \uparrow}$ and 
$\delta_{\downarrow}=\omega_{1 \, \downarrow}- \omega_{2 \, \downarrow}$,  $\omega_{1,2 \, \uparrow,\downarrow}=\alpha \gamma +\mu(\frac{P_{1,2 \, \uparrow,\downarrow}}{\gamma}-1)$ are free-running frequencies of the polarizations.
The dynamics of these phases is governed by the equations
\begin{eqnarray}
\dot \phi=\delta - 2 J_s \sin(\phi) - G_s \sin (\phi_{\uparrow}) + G_c \cos (\phi_{\uparrow}) + G_s \sin (\phi_{\downarrow}) - G_c \cos (\phi_{\downarrow})  \label{ph_equt_1} \\
\dot \phi_{\uparrow}=\delta_{\uparrow}     - 2  G_s \sin (\phi_{\uparrow})   - J_s \sin(\phi)+ J_c \cos(\phi)  \label{ph_equt_2} \\
\dot \phi_{\downarrow}=\delta_{\downarrow} - 2  G_s \sin (\phi_{\downarrow}) + J_s \sin(\phi)+ J_c \cos(\phi)  \label{ph_equt_3},
    \end{eqnarray}
where the coupling coefficients are $J_s=\sigma \left(1+\alpha^2-\frac{\alpha \mu P_0}{\gamma^2} \right)$ ,   $J_c=\frac{\sigma \mu P_0}{\gamma^2}$, $G_s= \left(1+q\cos (2\Omega t)\right) \left( \epsilon_0 - \tilde \epsilon_0 \left(\frac{\mu P_0}{\gamma^2}-\alpha \right) \right)$ and 
$G_c=\left(1+q\cos (2\Omega t)\right) \left( \tilde \epsilon_0 + \epsilon_0 \left(\frac{\mu P_0}{\gamma^2}-\alpha \right) \right) $. The fact that the system can be characterized by only three phases is the consequence of the invariance of the initial equations to the phase shift of all fields $A \rightarrow Ae^{i\theta}$. 
Let us note that the noise can be accounted by assuming that the pump fluctuation causes the fluctuations of the free-running frequencies of the modes and thus $\delta$, $ \delta_{\uparrow}$ and $\delta_{\downarrow}$ are fluctuating parameters.

Now we can also find the applicability limit of our approach. Above we assume that the evolutions of the phases are slow compared to the relaxation rate $\Gamma$. According to (\ref{ph_equt_1})-(\ref{ph_equt_3}) the fastest evolution rate of the phases $\phi$, $\phi_{\uparrow}$ and $\phi_{\downarrow}$ can be estimated as $\max \{ \delta, \delta_{\uparrow}, \delta_{\downarrow}, 2\Omega  \}$. In its turn, from (\ref{eq_rho_cL_pUp})-(\ref{eq_rho_cR_pDown}) it follows that the variation rate of $r$ are defined by the variation rates of these phases. Thus our approximation works provided that $\Gamma \gg \max \{ \delta, \delta_{\uparrow}, \delta_{\downarrow}, \Omega  \}$. 

It is also necessary to provide that the variation of the mode amplitudes are small compared to their unperturbed values, $r \ll \bar \rho$. To satisfy this it is sufficient to require that $\max \{ \sigma, \alpha \sigma, \epsilon_0, \tilde \epsilon_0 \} \ll \Gamma$.

In this paper we are interested in the dynamics in the range of detunings comparable to the coupling of the modes, so we describe the system correctly provided that the coupling between the modes and the pump rotation velocity are small compared to the relaxation rate of the condensate densities $\Gamma$. Our experimental parameters satisfy these conditions well.

Equations (\ref{ph_equt_1})-(\ref{ph_equt_3}) can easily be simulated numerically. To compare the results of numerical simulations with the experiments we need to express $g^{(2)}$ correlation function through the phases $\phi$, $\phi_{\uparrow}$ and $\phi_{\downarrow}$. Let us define the correlation function $g^{(2)}_{\theta_{pl}}$ of the signals measured from the $C_1$ or $C_2$ condensate through a linear polarizer oriented at the angle $\theta_{pl1}$ (for the first signal) and $\theta_{pl2}$ (for the second signal). This quantity is 
\begin{equation}
g^{(2)}_{\theta_{pl1}\theta_{pl2}}=\frac{ < I_{\theta_{pl1}}(t+\tau)I_{\theta_{pl2}}(t) > }{  < I_{\theta_{pl1}}(t) > < I_{\theta_{pl2}}(t) > } \label{auto_corr}
\end{equation}
where $I(t)$ is the temporal dependency of the intensity measured through the polarizer and $<>$ brackets denote averaging over time $t$. 

In the basis of circular polarization for the $C_1$ condensate we have 
\begin{equation}
I_{1\theta_{pl}}= \left| \frac{1}{\sqrt{2}}  \left( \rho_{1 \uparrow} e^{i\omega_{1 \uparrow}t+i\phi_{1 \uparrow}} e^{-i\theta_{pl}} +\rho_{1 \downarrow} e^{i\omega_{1 \downarrow}t+i\phi_{1 \downarrow}} e^{i\theta_{pl}} \right) \right|^2 =\frac{1}{2} \left(  |\rho_{1 \uparrow}|^2 +|\rho_{1 \downarrow}|^2 +2\rho_{1 \uparrow}\rho_{ 1 \downarrow}\cos((2\Omega t+\phi-2\theta_{pl})\right). \label{I_left_c}
\end{equation}

For the $C_2$ condensate the intensity measured through a linear polarizer oriented at the angle $\theta_{pl}$ is
\begin{eqnarray}
I_{2 \theta_{pl}}= \left| \frac{1}{\sqrt{2}}  \left( \rho_{2 \uparrow} e^{i\omega_{2\uparrow}t+i\phi_{2\uparrow}} e^{-i\theta_{pl}} + \rho_{2 \downarrow} e^{i\omega_{2\downarrow}t+i\phi_{2\downarrow}} e^{i\theta_{pl}} \right) \right|^2 =  \nonumber \\
=\frac{1}{2} \left(  |\rho_{2 \uparrow}|^2 +|\rho_{2 \downarrow}|^2 +2\rho_{2 \uparrow}\rho_{2 \downarrow}\cos( 2\Omega+\phi-\phi_{\uparrow}+\phi_{\downarrow} -2\theta_{pl})\right). \label{I_right_c}
\end{eqnarray}
Considering that in our approximation $\rho_{\uparrow} \approx \rho_{\downarrow}=\rho $ we obtain 
\begin{equation}
I_{1 \theta_{pl}}= \rho^2 \left(  1+\cos((2\Omega t+\phi -2\theta_{pl})\right)
\label{I_left_c_appr}
\end{equation}
for the $C_1$ condensate and 
\begin{equation}
I_{2 \theta_{pl}}= \rho^2 \left(  1+\cos((2\Omega t+\phi-\phi_{\uparrow}+\phi_{\downarrow} -2\theta_{pl})\right)
\label{I_right_c_}
\end{equation}
for the $C_2$ condensate.
This allows us to write the correlation function as 
\begin{equation}
g^{(2)}_{\theta_{pl1}\theta_{pl2}}=\frac{ < \left(1+ \cos(2\Omega (t-\tau) + \tilde \phi_{1}(t-\tau) -2\theta_{pl1}\right)\left(1+ \cos(2\Omega t + \tilde \phi_{2} (t)-2\theta_{pl2}\right)    >}{\left( 1+<\cos(2\Omega t + \tilde \phi_1(t)-2\theta_{pl1})> \right)\left( 1+<\cos(2\Omega t + \tilde \phi_2(t)-2\theta_{pl2})> \right)},
\label{corr_thr_phase_gen}
\end{equation}
where $\tilde \phi=\phi$ for the $C_1$ condensate and $\tilde \phi=\phi-\phi_{\uparrow}+\phi_{\downarrow}$ for the $C_2$ condensate, the index of $\tilde \phi$ shows whether the phase relates to the first or the second signal.

Let us remark that, as expected, the correlation function does not depend on the real amplitudes $\rho$. Calculating the values averaged over time $t$ one can change variables $t=t^{\prime}+\frac{\theta_{pl2}}{\Omega}$ and then it become evident that the correlation function depends only on the difference $\theta_{pl1}-\theta_{pl2}$ of polarizers orientation angles.
Let us also note
that in general the variations of the real amplitudes $r$ also affect the correlation function but  this contribution can be estimated as  $\left| \frac{r}{\rho} \right| \ll 1$ and thus can be neglected.

Now let us write down the expression for the auto- and  cross-correlation functions when both signals are measured through the linear polarizers oriented at the same angle $\theta_{pl}$. We have  
\begin{equation}
g^{(2)}_{\theta_{pl}}=\frac{ < \left(1+ \cos(2\Omega (t-\tau) + \phi(t-\tau) \right)\left(1+ \cos(2\Omega t + \phi(t) \right)    >}{\left( 1+<\cos(2\Omega t + \phi(t))> \right)^2 }
\label{autocorr_left}
\end{equation}
for the autocorrelation function measured from the $C_1$ condensate
\begin{equation}
g^{(2)}_{\theta_{pl}}=\frac{ < \left(1+ \cos(2\Omega (t-\tau) + \phi(t-\tau)-\phi_{\uparrow}(t-\tau)+\phi_{\downarrow}(t-\tau) \right)\left(1+ \cos(2\Omega t + \phi(t) -\phi_{\uparrow}(t)+\phi_{\downarrow}(t)\right)    >}{\left( 1+<\cos(2\Omega t + \phi(t)-\phi_{\uparrow}(t)+\phi_{\downarrow}(t))> \right)^2 }
\label{autocorr_right}
\end{equation}
for the autocorrelation function measured from the $C_2$ condensate and
\begin{equation}
g^{(2)}_{\theta_{pl}}=\frac{ < \left(1+ \cos(2\Omega (t-\tau) + \phi(t-\tau) \right)\left(1+ \cos(2\Omega t + \phi(t) -\phi_{\uparrow}(t)+\phi_{\downarrow}(t)\right)    >}{\left( 1+<\cos(2\Omega t + \phi(t))> \right)\left( 1+<\cos(2\Omega t + \phi(t)-\phi_{\uparrow}(t)+\phi_{\downarrow}(t))> \right) }
\label{cross_corr}
\end{equation}
for the cross-correlation function.

The simulations are done using the equations (\ref{ph_equt_1})-(\ref{ph_equt_3}) and then the required correlation functions are calculated using formulae (\ref{autocorr_left}),(\ref{autocorr_right}) or (\ref{cross_corr}). 
The coefficients in the equations (\ref{ph_equt_1})-(\ref{ph_equt_3}), except for the couplings, can be obtained by fitting the theoretical dependencies obtained to experimental measurements. To find the coupling between the polarization we can take the experimental data on the range of the pump rotation angular velocity where the polarization plane is locked to the rotating pump, see \cite{lxv5-qwxw} for more detail. It is more difficult to determine the coupling strength between the condensates. Following \cite{lagoudakis2017polariton,Berloff2017} we assume that the coupling is mostly dissipative and then fit the effective coupling strength which gives the best match between the theory and the experiment. We have checked that the accounting for the complex coupling does not affect the results qualitatively. Let us also remark that it is sufficient to fit the coupling strength for one series of the experiments and the found parameters allows to reproduce well the results of the other experiments.

\subsection{Synchronization of the modes and its effect on the correlation functions}

Let us first consider how the interaction between the modes affect the frequencies of the circular polarizations in the $C_1$ and in the $C_2$ condensates. We take two identical condensates where the free-running frequencies of the circular polarizations are equally in both condensates $\omega_{1 \uparrow}=\omega_{2 \uparrow}$, $\omega_{1 \downarrow} =\omega_{2 \downarrow}$. For our modelling we chose the detunings $\omega_{1,2 \uparrow}-\omega_{1,2 \downarrow}$ to be $ \approx 1.2$ GHz. The $C_1$ potential is driven by the rotating pump and we study how the detunings of the frequencies of the condensate modes depend on this parameter.

If the pump rotation velocity is low so that  the absolute value of the detuning $\delta=\omega_{1 \, \uparrow}- \omega_{1 \, \downarrow}-2\Omega$ is large compared to the couplings then, as it is seen from (\ref{ph_equt_1}), $\phi$ is a quasi-linear function of time rapidly growing at the rate $\omega_{1\uparrow}-\omega_{1\downarrow}$. This means that the polarization plane of the $C_1$ condensate  rotates at the velocity close to $\frac{1}{2}(\omega_{1 \uparrow}-\omega_{1 \downarrow})$.

In this case the last two terms in (\ref{ph_equt_2}) and (\ref{ph_equt_3}) are rapidly oscillating functions and its contribution to the slow dynamics of $\phi_{\uparrow}$ and $\phi_{\downarrow}$ can be neglected.  The phases are described by the 
\begin{eqnarray}
\dot \phi_{\uparrow}=\delta_{\uparrow}     - 2  G_s \sin (\phi_{\uparrow})    \label{ph_equt_2_m1} \\
\dot \phi_{\downarrow}=\delta_{\downarrow} - 2  G_s \sin (\phi_{\downarrow})   \label{ph_equt_3_m1}.
\end{eqnarray}
If the cross-condensates coupling does not depend on time then the equations (\ref{ph_equt_2_m1}),(\ref{ph_equt_3_m1}) have  equilibrium points being global attractors provided that $|\delta_{\uparrow}|<|G_s|$. Weak dependency of the coupling coefficient on time make the stationary solutions weakly oscillating around the equilibrium points. 
This means that the the clockwise  polarizations of the condensates are locked as well as the counter-clockwise ones. So the polarization planes of the condensate rotate at the same velocity and the average angle between them is fixed. 

\begin{figure}[t]
    \centering
    \includegraphics[width=0.9\columnwidth]{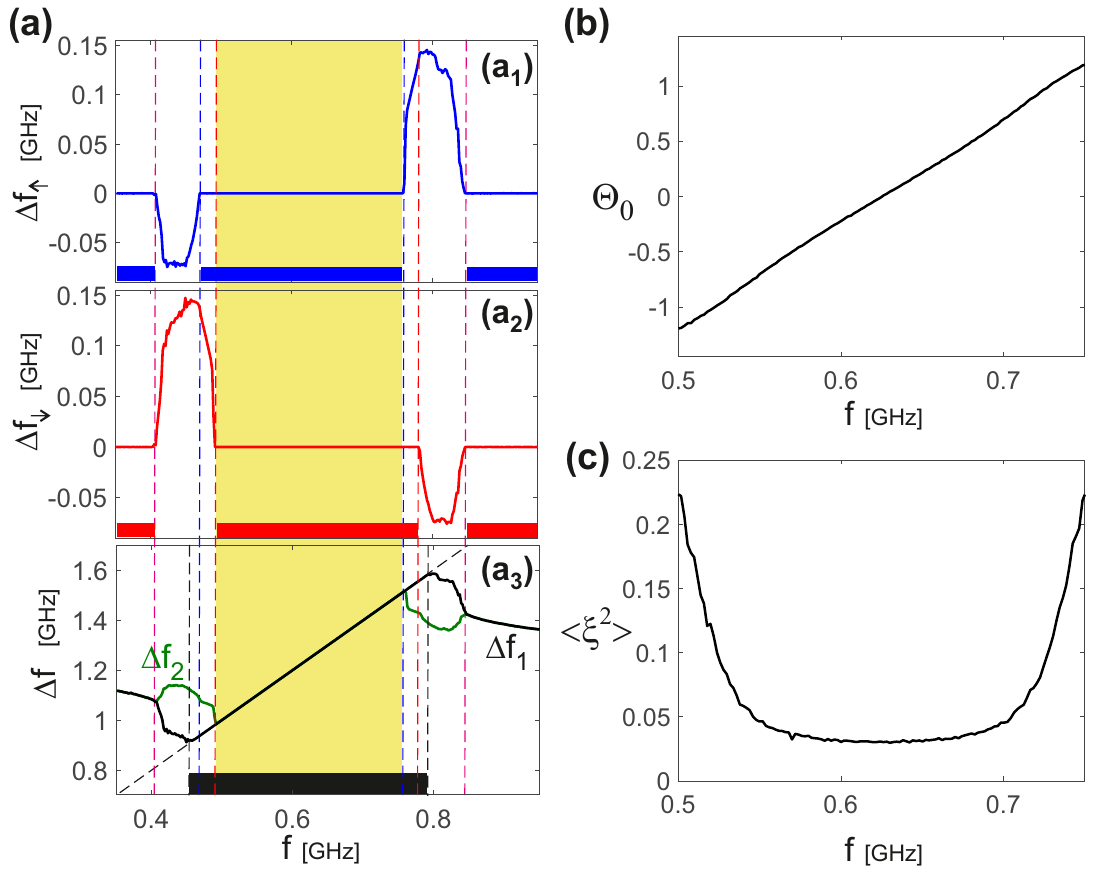}
    \caption{Panel (a) illustrates the synchronization in the case of the different polarizations of the condensates. The upper part (a$_1$) show the difference of the $\uparrow$ polarization frequencies in the $C_1$ and the $C_2$ condensates. The frequency difference is $\Delta \omega_{\uparrow} = \omega_{1\uparrow}-\omega_{2,\uparrow}+<\dot \phi_{1\uparrow}>-<\dot \phi_{2\uparrow}>$. 
     The medium part (a$_2$) shows the same but for $\downarrow$ polarization. The black curve in the bottom part (a$_3$) or the panel shows the frequency difference $\Delta \omega_1=\omega_{1\uparrow}-\omega_{1\downarrow}+<\dot \phi>$ of the $\uparrow$ and $\downarrow$ polarizations in the $C_1$ condensate. The green line shows the same but for the $C_2$ condensate, $\Delta \omega_2=\omega_{2\uparrow}-\omega_{2\downarrow}+<\dot \phi>-,\dot \phi_{\uparrow}+<\dot \xi_{\downarrow}>$. For convenience we measure the detunings as $\Delta f_{\uparrow, \downarrow}=\frac{\Delta \omega_{\uparrow, \downarrow}}{2\pi}$ (the vertical axis'), and the pump rotation velocity as $f=\frac{\Omega}{2\pi}$ ( the horizontal axis'). 
    The vertical dashed magenta lines indicate the thresholds where the synchronization between the polarizations in the $C_1$ and $C_2$ condensates is broken. The vertical dashed black lines indicate the thresholds at which the polarizations in the $C_1$ condensate become asynchronously locked due to the action of the rotating potential. The blue and red dashed lines show the thresholds at which the $\uparrow$ and $\downarrow$ polarizations in the $C_1$ and $C_2$ condensates become locked, respectively. The inclined black dashed line in (a$_3$) marks the frequency difference equal to twice the pump rotation velocity.  The blue bars mark the intervals of the pump rotation velocities where the $\uparrow$ polarizations in the $C_1$ and the $C_2$ condensates are synchronized. The red bars show the same but for $\downarrow$ polarizations. The black bar shows the interval where the $\uparrow$ and $\downarrow$ polarizations in the $C_1$ condensate are asynchronously phase locked. The yellow color highlights the interval of pump rotation velocities where the fully synchronized regime occurs. Panel (b) shows the dependence of the angle $\Delta \Theta_0$ between the polarization planes in the $C_1$ and $C_2$ condensates on the pump rotation velocity. The dependence of the phase dispersion $\langle \xi^2 \rangle$ on the pump rotation velocity is presented in panel (c).   } 
    \label{fig_S_synchr}
\end{figure}

Let us illustrate it by the results of numerical simulations. In Fig.~\ref{fig_S_synchr}(a$_1$)(a$_2$) the frequency differences $\Delta \omega_{\uparrow,\downarrow} = \omega_{1\uparrow,\downarrow}-\omega_{2\uparrow,\downarrow}+<\dot \phi_{1\uparrow,\downarrow}>-<\dot \phi_{2\uparrow,\downarrow}>$ obtained in the numerical simulations are shown as functions of the pump rotating velocity. It is seen that there are intervals where the frequencies becomes equal, in particular the synchronization occurs for the low and high pump rotation velocities.  We have checked that in these interval the mutual phase between the $\uparrow$ polarizations in the $C_1$ and the $C_2$ condensates is fixed and does not depend on the initial conditions. The noise results in fluctuation of the mutual phase around its mean value. The same applies to $\downarrow$ polarization.

In the intervals corresponding to the low and high pump rotation velocity the frequency difference $\Delta \omega_{1, 2}=\omega_{1,2 \uparrow}-\omega_{1,2\downarrow}+<\dot \phi_{1,2 \uparrow}>-<\dot \phi_{1,2\downarrow}>$ between the $\uparrow$ and $\downarrow$ polarizations in both condensates does not coincide with the doubled rotation velocity of the pump, see Fig.~\ref{fig_S_synchr}(a$_3$). So the condensates polarization planes rotate at the velocity different from $\Omega$, $\Delta \omega_{1, 2} \neq 2 \Omega$.  

In the same the $\uparrow$ as well as  $\downarrow$ polarizations in the $C_1$ and $C_2$ condensates are pairwise synchronized, $\Delta \omega_{1}=\Delta \omega_{2}$. This synchronization means that the polarization planes in both condensates rotate at exactly the same velocity, see Fig.~\ref{fig_S_synchr}(a$_3$),   and so the mean angle between the polarization plane is fixed.  It worth noting that the locking between the polarization planes does not suppress their fluctuation but make them fluctuating in unison so that the angle between the polarization planes is constant.

Shifting the the pump rotation velocity $\Omega$ closer to the resonant one defined as $\Omega_{r}=\frac{\omega_{1\uparrow}-\omega_{1\downarrow}}{2}$ makes the impact of pump rotation stronger and at some point this destroys the synchronization,  see Fig.~\ref{fig_S_synchr}(a$_1$)(a$_2$) showing that the polarization frequencies detuning become different in the $C_1$ and $C_2$ condensates, Fig.~\ref{fig_S_synchr}(a$_3$). This, in its turn, results in the rotation of the polarization planes in the $C_1$ and $C_2$ condensates at different velocities. 

At some critical pump rotation velocity the interaction between the polarizations in the $C_1$ condensate becomes strong enough for the polarizations to get synchronized. The synchronization implies that the frequency difference between the polarizations in the $C_1$ condensates is exactly equal to $2\Omega$, see Fig.~\ref{fig_S_synchr}(a$_3$). In the $C_2$ condensate the polarization plane rotates at a different velocity. 

In the regime of synchronous rotation of the polarization plane in the $C_1$ condensate the frequency of both polarizations in this condensate depends on $\Omega$. For the chosen parameters the drawing of $\Omega$ closer to $\Omega_r$ causes such shift of the polarization frequencies that the detunings  $\Delta \omega_{\uparrow}$ and $\Delta \omega_{\downarrow}$ decrease. Then first one (in our case $\uparrow$) then another (in our case $\downarrow$) polarization in the $C_1$ condensate gets synchronized to its counterpart in the $C_2$ condensate. Then, as it is seen in Fig.~\ref{fig_S_synchr}(a$_1$)(a$_2$), the frequencies of the $\uparrow$ polarization in both condensates become the same, the frequencies of and $\downarrow$ polarization become the same too. 

The difference between the $\uparrow$ and $\downarrow$ polarizations in these fully synchronized regime is equal exactly to $2\Omega$, Fig.~\ref{fig_S_synchr}(a$_3$). 
Let us remark that the existence of such regime is obvious from (\ref{ph_equt_1})-(\ref{ph_equt_3}) for $\delta=0$, $\delta_{\uparrow,\downarrow}=-J_c \frac{J_s}{|J_s|}$, the exact borders of the synchronization in the parameter space can be found by rigorous analysis of (\ref{ph_equt_1})-(\ref{ph_equt_3}).

In the fully synchronized regime the polarization planes in both condensate rotate synchronously with the pump and the fluctuations of both polarization planes are suppressed. Let us consider it in more detail. First, let us define the mean angle between the polarization planes as 
\begin{equation}
\Theta_0=-\frac{i}{2}\log \left( \frac{<\exp(2i\theta)>}{ z } \right)  \label{mean_angle_def}
\end{equation}
where $\theta$ is the fluctuating value of the angle between the polarization planes and $z=\left|<\exp(2i\theta))>\right|$.
Let us remark that this definition is equivalent to the equality
\begin{equation}
e^{2i\Theta_0}=\frac{<e^{2i\theta}>}{\left| <e^{2i\theta}> \right| } \label{mean_angle_def_equiv}
\end{equation}
% It is easy to check that the shift of the angle $\theta$ by some value $\Theta_{\Delta}$, $\tilde \theta=\theta+\Theta_{\Delta}$, shifts the mean angle by the same value, $\tilde \Theta_0= \Theta_0 +\Theta_{\Delta} $.

Let us note that, in general, $|\exp(2i\theta)| \neq 1$. For example, if the polarization planes rotate at constant but different velocities then the angle increase linearly in time and  $|<\exp(i\theta)>| =0 $. It is obvious that in this case the mean angle does not have much physical sense.  However, if most of the time the angle between the polarizations is close to some value, then the mean angle $\Theta_0$ can be used to characterize the dynamics. It is easy to see that if $\theta$ is a constant equal to $\theta_0$ (no fluctuations) then $\Theta_0=\theta_0$. If the fluctuations of the angle around its mean value are weak then $|<\exp(i\theta)>| \approx 1$.

The dependency of the angle between the polarization planes in the $C_1$ and $C_2$ condensates is shown in Fig.~\ref{fig_S_synchr}(b) for the fully synchronized regime. The dependency does not depend on the initial condition and confirms the mutual phase locking of the polarization planes. Considering that the polarzation plane of the $C_1$ condensate is locked to the rotating pump we can conclude that in this regimes both polarization planes are locked to the rotating pump and the fluctuations are suppressed. This way the coherence of the rotation of the polarization plane is transferred from the $C_1$ to the $C_2$ condensate. 

Let us demonstrate the suppression of the fluctuation by calculating the mean deviation of the angle from its mean value. This can be characterized by the quantity $z$ introduced above. Let us reveal the physical meaning of $z$ in the case of weak fluctuation of the angle between the polarization planes. Then we can represent the instantaneous angle  as $\theta=\Theta_0+\xi$ where $\xi \ll 1$ is a small deviation of the angle from its mean value. Then we can calculate $<\exp(i\theta)>$ expanding it in Taylor series and keeping three leading terms $<\exp(2i\theta)>\approx \exp(2i\Theta_0)<1+2i\xi +2 \xi^2 >$. 

Using the definition of the mean angle (\ref{mean_angle_def_equiv}) we obtain that 
$<1+2i\xi - 2 \xi^2 >= z$. By definition $z$ is positive, therefore $<\xi>=0$. For the mean squared deviation we obtain $<\xi^2>=\frac{1-z}{2}$.

The numerically calculated mean square $<\xi^2>$ is shown in Fig.~\ref{fig_S_synchr}(c). As expected, close to the synchronization edges the effect of the fluctuations becomes more pronounced. If the pump rotation velocity is close to $\Omega_r$ then the deviation of the angle between the polarization plane is very small and we can speak about coherently rotating polarization planes. 
Let us remark that the angle between the polarization planes as well as the mean squared deviation can be found in experiment from the measurements of the correlation functions. This will be considered below.

\begin{figure}[h]
    \centering
    \includegraphics[width=0.7\columnwidth]{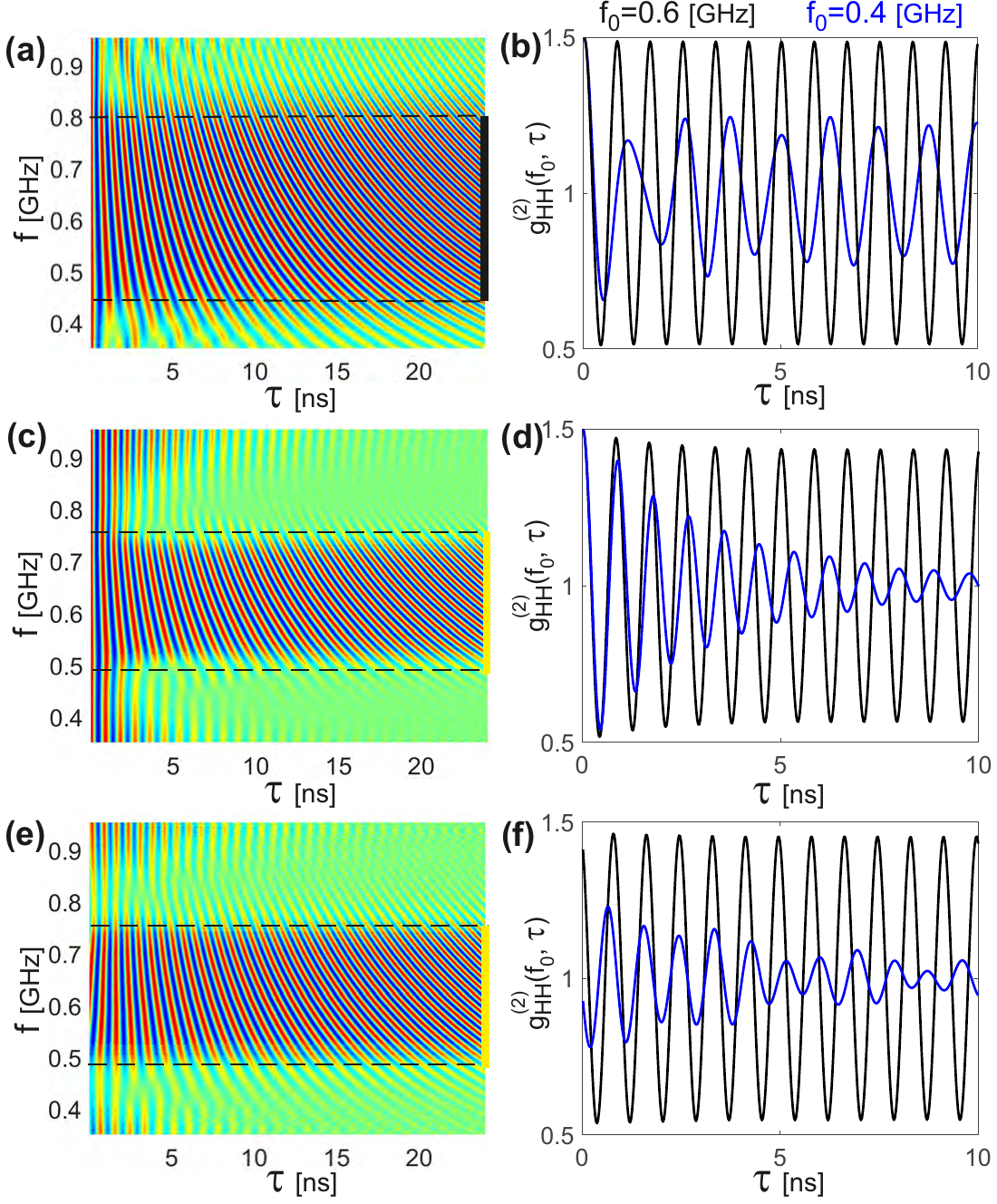}
    \caption{The dependence of the correlation function $g^{(2)}$ on the pump rotation velocity for signals measured from the driven condensate through a horizontally oriented polarizer is shown in panel (a). 
    The dashed horizontal line and the vertical bar in the right part of the panel show the pump rotation velocity range where the polarization plane of the $C_1$ condensate is phase-locked to the pump.     
    The correlation functions for the synchronous ($\Omega = 0.6$ GHz) and non-synchronous ($\Omega = 400$ GHz) regimes are presented in panel (b).
    Panels (c) and (d) show the same quantities but for the $C_2$ (non-driven) condensate. 
    Panels (e) and (f) display the cross-correlation functions of the signals measured from the $C_1$ and $C_2$ condensates through horizontally oriented polarizers.
    The dashed lines and the yellow vertical bar in panels (c) and (e) show the range of the pump rotation velocities where fully synchronized regime occurs.  The vertical axis's in panels (a),(c) and (e) are in units of $f=\frac{\Omega}{2\pi}$.  } 
    \label{fig_correlations_co_dir}
\end{figure}

In our experiments the measurable quantities are the auto and cross-correlation functions. Let us discuss how the synchronization manifests itself in the behavior of the correlation functions $g^{2}$. In Fig.~\ref{fig_correlations_co_dir} panel (a) show the correlation function of the signals measured from the $C_1$ condensate through the same linear polarizer.  It is seen that there is a range of the pump rotation velocity where the correlation function oscillations slightly decrease and then reach a very long plateau lasting for hundred of nanoseconds. This happens exactly in the range where the polarization plane of the $C_1$ condensate is locked to the rotating pump. If the rotation velocity lies outside the synchronization interval then the correlation function oscillations rapidly decay to  zero. This is illustrated in panel (b) showing the autocorrelation functions for $\Omega=0.6$ GHz (withing the synchronization region) and for $\Omega=0.4$ GHz (out of the synchronization region). This, actually, reproduces the results of \cite{gnusov2024observation,lxv5-qwxw} where it is shown that the locking of the polarization plane to the rotating pump allows to obtain coherent rotation of the polarization plane.

The same correlation functions but measured from the $C_2$ condensate is shown in panels (c) and (d). It is seen, that in spite of the fact that this condensate is not directly affected by the rotating pump there is a range of pump rotation velocities where the correlation function exhibit long lasting oscillations indicating that the rotation of the polarization plane is synchronized with the rotating pump in the neighborliness condensate. Additional confirmation of the synchronization is the observation that the frequency of the correlation function oscillation is equal exactly to $2\Omega$. 

The existence of the cross-correlations between the polarization planes in the $C_1$ and the $C_2$ condensates is shown in Fig.~\ref{fig_correlations_co_dir}(e) showing the correlation function of a signal from the $C_1$ and a signal from the $C_2$ condensates. Both signal are measured through the co-aligned linear polarizers. It is seen, that there is a range of pump rotation velocities where oscillations of the cross-correlation  function last for long time. The correlation functions calculated for different $\Omega$ and demonstrating different behavior are shown in Fig.~\ref{fig_correlations_co_dir} panel (f).

\begin{figure}[t]
    \centering
    \includegraphics[width=0.9\columnwidth]{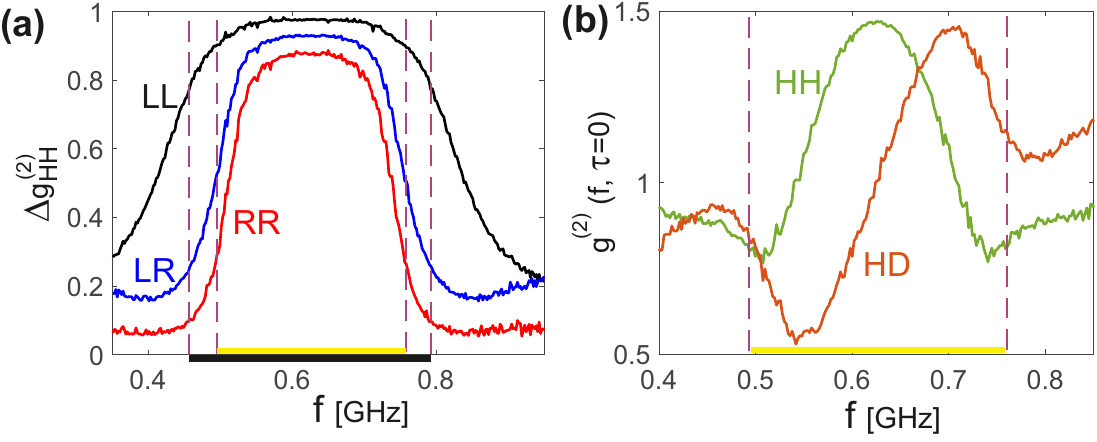}
    \caption{Panel (a) shows the dependence of the oscillation amplitudes of the correlation functions at long times on the pump rotation velocity. The black curve marked as LL is the for the auto-correlation function of the signal measured from the $C_1$ condensate through a horizontal linear polarizer. The red curve marked as $RR$ is the same but for the signal measured from the $C_2$ condensate. The blue curve marked as LR is for the cross-correlation function of the signals measured from the $C_1$ and $C_2$ condensates through the horizontal polarizer.  
The dependence of the cross-correlation function at $\tau = 0$ on the pump rotation velocity is shown in panel (b). The green curve (HH) corresponds to the case where the signals from both condensates are measured through horizontally oriented polarizers. The red curve (HD) represents the correlation function between the signal from the $C_1$ condensate measured through a horizontal polarizer and the signal from the $C_2$ condensate measured through a polarizer oriented at $\frac{\pi}{4}$ to the horizontal axis. 
    The black horizontal bar shows the interval of the pump rotation velocities where the polarization plane of the $C_1$ condensate is locked to the rotating pump. The yellow bars show the range of the rotation velocities where fully synchronized regime occurs. The vertical dashed lines show the borders of the regions. For sake of convenience the along the horizontal axis' are in units of the quantity $f=\frac{\Omega}{2\pi}$.
      } 
    \label{fig_correlations_ampl}
\end{figure}

To estimate the ranges of $\Omega$ where the amplitudes of the correlation functions oscillations last for long time we calculated the dependency of the oscillations amplitude at relatively long time corresponding to the plateau there the amplitudes stays nearly constant. These dependencies are shown in  Fig.~\ref{fig_correlations_ampl}(a). The auto-correlation function calculated for the $C_1$ condensate has the wide region of $\Omega$ where the long lasting oscillations are pronounced. It is also seen that this range approximately coincides with the synchronization range of the polarization plane of the $C_1$ condensate with the rotating pump.

The autocorrelation function of the $C_2$ condensate and the cross correlation function have pronounced long lasting oscillations in a more narrow ranges of the rotation velocities $\Omega$. These ranges are not identical but they are close to each other and approximately coincide with the range of $\Omega$ where fully synchronized regime occurs, see Fig.~\ref{fig_correlations_ampl}(a). 

Let us remark that the correlation function allows us to find the angle between the polarization planes of the condensates in the regime of full synchronization. This can be usefull from the experimental point of view considering that it is not possible to measure directly the mean value of $\theta$.

To do this all we need is the information on the cross-correlation functions measured for zero delay time $\tau=0$ for the signals coming through the polarizers aligned at different angles, for example through co-aligned polarizers and through polarizers aligned at the angle $\frac{\pi}{4}$. These dependencies can be found experimentally or obtained from the numerical simulations, see Fig.~\ref{fig_correlations_ampl}(b). Below we show that this information is sufficient to find the mean angle between the polarization planes and the mean squared deviation of the angle. 

To proceed we study how the cross-correlation function (\ref{corr_thr_phase_gen}) depends on $\Omega$ setting the delay time to zero $\tau=0$. This value is given by 
\begin{equation}
g^{(2)}_{\theta_{pl1}\theta_{pl2}}=\frac{ < \left(1+ \cos(2\Omega t +\phi(t) -2\theta_{pl1})\right)\left(1+ \cos(2\Omega t + \phi(t)-\phi_{\uparrow}(t)+\phi_{\downarrow}(t) -2\theta_{pl2})\right)    >}{\left( 1+<\cos(2\Omega t + \phi(t)-2\theta_{pl1})> \right)\left( 1+<\cos(2\Omega t + \phi(t)-\phi_{\uparrow}(t)+\phi_{\downarrow}(t) -2\theta_{pl2})> \right)}.
\label{corr_corr_synchr}
\end{equation}

For weak fluctuation the evolution of the phase $\phi$ is slow compared to the rotation velocity $\Omega$. Thus the term $<\cos(2\Omega t +  \phi(t) -2\theta_{pl1})>=<\cos(2\Omega t-2\theta_{pl1})\cos( \phi )>-<\sin(2\Omega t-2\theta_{pl1})\sin( \phi )> \ll 1$ because of the averaging of the product of the slow function of time of the frequently oscillating $\cos$ ($\sin$) function. 
The same reasoning shows that $|<\cos(2\Omega t + \phi-\phi_{\uparrow}+\phi_{\downarrow} -2\theta_{pl2})>| \ll 1$. This allow to rewrite (\ref{corr_corr_synchr}) as
\begin{eqnarray}
g^{(2)}_{\theta_{pl1}\theta_{pl2}}=1 + < \cos(2\Omega t + \phi -2\theta_{pl1}) \cos(2\Omega t + \phi-\phi_{\uparrow}+\phi_{\downarrow} -2\theta_{pl2})    > = \nonumber \\
=1+\frac{1}{2} < \cos(\phi_{\uparrow}-\phi_{\downarrow} -2(\theta_{pl1}-\theta_{pl2}) )    >
+\frac{1}{2} <  \cos(4\Omega t + 2\phi-\phi_{\uparrow}+\phi_{\downarrow} -2(\theta_{pl1}+\theta_{pl2}) ))    >.
\label{corr_corr_synchr2}
\end{eqnarray}

The last term in (\ref{corr_corr_synchr2}) is small in the case of weak fluctuations and can be neglected, see the reasoning above. The angle between the polarization planes in the $C_1$ and $C_2$ condensates can be expressed through the phases $\phi_{\uparrow}$, $\phi_{\downarrow}$ as $\theta=\frac{\phi_{\uparrow}-\phi_{\downarrow}}{2}$. Then we have the expression for the cross-correlation function at $\tau=0$ in the form 
\begin{eqnarray}
g^{(2)}_{\theta_{pl1}\theta_{pl2}}=1+\frac{1}{2} < \cos(2(\theta_{pl1}-\theta_{pl2}-\theta) )    >
\label{corr_corr_synchr3}
\end{eqnarray}

The angle between the polarization planes in the case of weak fluctuations can be found from the cross-correlation function $g^{(2)}$.
Knowing the correlation function of the signals measured through co-aligned polarizers $g^{(2)}_{HH}$ it is straightforward to express the average $<\cos(2\theta)>$ through the correlation function $<\cos(2\theta)>=2\left( g^{(2)}_{HH}-1 \right)$. 
Representing the angle between the polarization as $\theta=\Theta_0 +\xi$  we expend the cosine function into Taylor series up to the second order and obtain 
\begin{equation}
<\cos(2(\Theta_0+\xi))>=(1-2<\xi^2>)\cos(2 \Theta_0)=z\cos(2 \Theta_0)=2(g^{(2)}_{HH}-1). \label{cos_thet_corr}
\end{equation}
Analogously it can easily be shown that
\begin{equation}
<\sin(2(\Theta_0+\xi))>=(1-2<\xi^2>)\sin(2 \Theta_0)=z\sin(2 \Theta_0) =2(g^{(2)}_{HD}-1), \label{sin_thet_corr}
\end{equation}
where $g^{(2)}_{HD}$ is the cross-corelation function measured through  the polarizers aligned at the angle $\frac{\pi}{4}$.

From here after a simple trigonometry it follows that 
\begin{equation}
z= 2\sqrt{\left(g^{(2)}_{HH}-1\right)^2+\left(g^{(2)}_{HD}-1\right)^2}.   \label{z_through_corr}
\end{equation}
The the mean angle can be found either from (\ref{cos_thet_corr}) or (\ref{sin_thet_corr}) as
\begin{equation}
\Theta_0=-\frac{i}{2} \log\left( \frac{\left(g^{(2)}_{HH}-1\right) +i\left(g^{(2)}_{HD}-1\right)}{\sqrt{\left(g^{(2)}_{HH}-1\right)^2+\left(g^{(2)}_{HD}-1\right)^2}}  \right)  \label{mean_angle}
\end{equation} 

the mean squared deviation is
\begin{equation}
<\xi^2>=\frac{1}{2}-\sqrt{\left(g^{(2)}_{HH}-1\right)^2+\left(g^{(2)}_{HD}-1\right)^2}. \label{mean_deviat}
\end{equation}